\documentstyle[preprint,aps,eqsecnum,psfig]{revtex}

\newcommand{\be}{\begin{equation}}
\newcommand{\ee}{\end{equation}}

\newcommand{\bea}{\begin{eqnarray}}
\newcommand{\eea}{\end{eqnarray}}

\newcommand{\ba}{\begin{array}}
\newcommand{\ea}{\end{array}}

\newcommand{\rr}{\right}
\newcommand{\lt}{\left}

\newcommand{\nn}{\nonumber}

\newcommand{\ds}{\displaystyle}

\begin{document}

\setlength{\unitlength}{1cm}

\draft

\title{Monte Carlo study of the growth of\\ $L1_2$ ordered domains
in fcc $A_3B$ binary alloys}

\author{Carlos Frontera, Eduard Vives, Teresa Cast\'an and Antoni Planes}

\address{Departament d'Estructura i Constituents de la Mat\`eria, \\ Facultat
de F\'{\i}sica, Universitat de Barcelona,\\ Avd. Diagonal 647, E-08028
Barcelona, Catalonia, Spain. \\ Tel.: 34 3 4021586, Fax: 34 3 4021174,
e-mail: carlos@ecm.ub.es}

\date{\today}

\maketitle

\begin{abstract}

A Monte Carlo study of the late time growth of $L1_2$ ordered domains on a
fcc $A_3B$ binary alloy is presented. The energy of the alloy has been
modeled by a nearest neighbor interaction Ising hamiltonian. The system
exhibits a fourfold degenerated ground-state and two kinds of interfaces
separating ordered domains: flat and curved antiphase boundaries. Two
different dynamics are used in the simulations: the standard atom-atom
exchange mechanism and the more realistic vacancy-atom exchange mechanism.
The results obtained by both methods are compared. In particular we study the
time evolution of the excess energy, the structure factor and the mean
distance between walls. In the case of atom-atom exchange mechanism
anisotropic growth has been found: two characteristic lengths are needed in
order to describe the evolution. Contrarily, with the vacancy-atom exchange
mechanism scaling with a single length holds. Results are contrasted with
existing experiments in $Cu_3Au$ and theories for anisotropic growth.

\end{abstract}


\narrowtext

\section{Introduction}
\label{Introduction}

Kinetics of phase transitions is a problem of great interest not only
because of its fundamental importance in non-equilibrium Statistical
Physics, but also because its many implications in different areas of
Material Science and Technology \cite{Gunton83,Komura88}. The phenomenon
is a consequence of the far from-equilibrium initial conditions induced by
the sudden change of the imposed thermodynamic parameters on time scales much
shorter than the time scales characterizing the process towards the new
equilibrium situation. Typically, the system is quenched through
its equilibrium ordering temperature. Immediately after the quench, domains
of the new phase appear. As time goes on, they grow in size in order to
reduce the excess free energy of the walls. This growth shows distinct
regimes from early to late times. At late times, in the so called domain
growth regime, it is usually assumed that the domain size is much larger than
all microscopic lengths in the system. Then, inspired by the situation in
equilibrium critical phenomena, it is assumed that the system shows dynamical
scaling \cite{Mazenko91}. This means that at different times the domain
structure looks the same when lengths are measured in units of the
characteristic length. Furthermore, the average domain size is supposed to
increase with time according to a power law with an exponent $x$
characteristic of the universality class to which the system is supposed to
belong \cite{Mouritsen90}.  If these universality classes really exist, the
important point is to identify the distinctive features of a given class. In
the simplest case, they are supposed to depend only on whether or not the
order parameter is conserved. A typical example for the conserved case is a
phase separation process while an order-disorder transition in a binary alloy
corresponds to the non-conserved situation. In the first case $x= 1/3$ has
been predicted while $x= 1/2$ is the expected value in the second case
\cite{Gunton83}. Nevertheless, the concept of universality is not firmly
established and is still under discussion. The above classification, just
based on the order parameter conservation property, seems to apply only when
no disorder is present in the system and the excitations are homogeneously
distributed. For instance it is well acknowledged that quenched disorder
gives rise to logarithmic growth laws \cite{Huse85}, and that excitations
localized on the interfaces give rise to exponents greater than 1/2 and 1/3
for the non-conserved and conserved cases respectively
\cite{Vives92,Frontera96}. Moreover other parameters like the ground state
degeneracy, non-stoichiometry or anisotropic effects could modify the
exponents. The exponent $x=1/2$ in non-conserved order parameter systems is
theoretically based on a curvature driven interface motion \cite{Allen79}.
This is the well known Allen-Cahn law for domain growth. A special curvature
driven case leading to an exponent $x=1/4$ has been found for anisotropic
systems with a mixture of perfectly flat and curved domain walls
\cite{Castan89}.

In this paper we will focus on the domain growth problem in a fcc $A_3B$
binary alloy undergoing an order-disorder transition from a disordered to a
$L1_2$ structure. This has been the system most widely used to perform
experiments intended to study ordering kinetics in non-conserved order
parameter systems
\cite{Marcinkowski61,Poquette69,Sakai71,Rase75,Bley76,Hashimoto78,Katano88,
Nagler88,Shannon92}. In spite of that, theoretical studies of domain growth
in fcc systems are really scarce and, as far as we know, only a recent paper
by Lai \cite{Lai90} is specifically devoted to the study of such kind of
systems. Concerning computer simulation no data has, to our knowledge, been
published. This is a rather surprising fact taking into account that Monte
Carlo numerical studies have been of seminal importance in providing much of
the quantitative insight into ordering kinetic problems. This is probably due
to the inherent complexity of ordering problems in fcc lattices. Actually the
ground state is fourfold degenerate and there exist two different kinds of
antiphase boundaries: high and low excess energy boundaries. Then, two
characteristic lengths may grow following different laws (anisotropic growth)
and this can, in some way, question the validity of scaling properties in
such a system. In fact, related problems have been considered in $2d$
lattices. For example, the fcc problem has some similarities with the problem
of island growth in a system of hydrogen adsorbed on a $(110)$ iron surface
with coverage 2/3 \cite{Vinals85}. In this case it has been found an exponent
smaller than 1/2 which is associated to interface diffusion effects. Also
some indications on anisotropic growth are reported in the same reference
\cite{Vinals85}. Concerning the experimental situation, among a rich
literature, it is worth mentioning a pioneer x-ray diffraction work by
Cowley \cite{Cowley50} devoted to the study of the superstructure peak in
$Cu_3Au$ single crystals close to equilibrium. Much more recently, a very
complete time-resolved x-ray scattering study in a single crystal of $Cu_3Au$
\cite{Shannon92}, has been published. The authors obtain $x=1/2$ with a high
degree of accuracy for the growth of the high energetic boundaries and with
less degree for the low energetic ones. This leads them to the conclusion
that scaling holds in this anisotropic system in agreement with the
theoretical predictions by Lai \cite{Lai90}.

Our interest in this paper is to present extensive Monte Carlo simulations of
the domain growth process performed on a fcc $A_3B$ alloy. Since we modelize
the system assuming pairwise interactions between nearest neighbor atoms
only, the low-energy boundaries have exactly zero excess energy. We expect
that this extreme situation emphasize any tendency of the system to show
anisotropic growth effects. In addition we will perform simulations in
systems either with and without vacancies (in this last case only
vacancy-atom exchanges will be allowed) in order to investigate the effect
of the vacancy mechanism in the kinetics of ordering in fcc lattices. This
mechanism has, recently, been shown to play a very important role in bcc
lattices \cite{Frontera94}.

The paper is organized as follows. In the next section we introduce the model
and describe some of the features of the ground state and the domain walls.
In section \ref{MCDetails} we explain the details of the Monte Carlo
simulations and define the different magnitudes used to describe the
ordering process. The results are sequentially presented in section
\ref{Results} and discussed in section \ref{Discussion}. Finally, in section
\ref{Summary}, we give a summary of the main conclusions.

\section{Model and ground state}
\label{MODEL}

The binary alloy is modeled by a set of $N_A$ $A$-atoms, $N_B$ $B$-atoms and
$N_V$ vacancies on a ``perfect'' fcc lattice with lattice spacing $a$, linear
size $a\,L$ and periodic boundary conditions. The number of lattice sites is
$N=4L^3= N_A+N_B+N_V$. Assuming nearest neighbors (n.n.) interactions only, a
general ABV model hamiltonian \cite{Yaldram91} can be rewritten as a
Blume-Emery-Griffiths \cite{Blume71} one, as explained in
Refs.~\cite{Frontera94,Vives93b}:

\be
{\cal H}= J \sum_{\lt \langle i,j\rr \rangle}^{\rm n.n.}S_i S_j+
   K \sum_{\lt \langle i,j\rr \rangle}^{\rm n.n.}S_i^2S_j^2+
   L \sum_{\lt \langle i,j\rr \rangle}^{\rm n.n.}\lt (S_i^2S_j+S_iS_j^2\rr
)+ {\cal H}_0 ,
\label{hbeg}
\ee
where the sums extend over all n.n. pairs and $i$ and $j$  are generic
indexes sweeping all the lattice ($i,j=1,\dots N$). The spin variables $S_i$
can take three values: $+1$, $-1$ and $0$ when the $i$-th position of the
lattice is occupied by an $A$-atom, a $B$-atom or a vacancy respectively. The
parameters $J$, $K$ and $L$ are coupling parameters and ${\cal H}_0$ is a
function of the concentration. In the case of low vacancy concentrations
($N_V\ll N_A, N_B$) this hamiltonian can be approximated by a spin-1 Ising
model:

\be
{\cal H}= J \sum_{\lt \langle i,j \rr \rangle}^{\rm n.n.} S_i S_j ,
\label{ham}
\ee
except for an irrelevant additive constant. This is the hamiltonian that we
have used in the present simulations.

\begin{figure}

\centerline{\hbox{
\psfig{figure=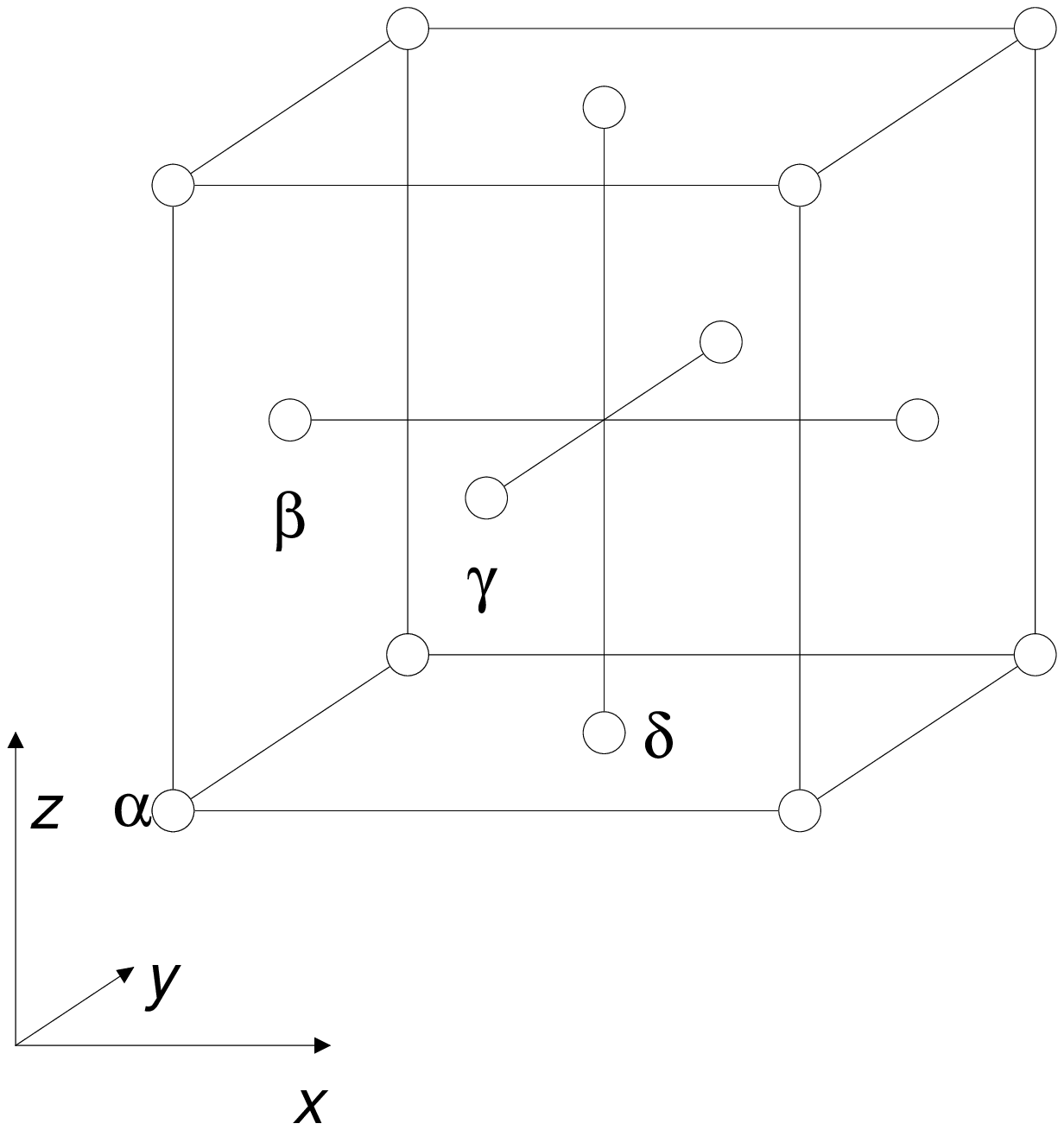,height=315pt,width=315pt}} }

\caption{The four sublattices $\alpha$, $\beta$, $\gamma$ and $\delta$ in
to which the fcc lattice can be divided.}

\label{sublatt}
\end{figure}

We have focused on the case $N_A \simeq 3 N_B \gg N_V$ in order to simulate
an $A_3B$ alloy like $Cu_3Au$, $Ni_3Mn$ or $Ni_3Fe$, with little
concentration of vacancies. It is well known that for $J>0$ and no vacancies
($N_V=0$) this system presents a discontinuous order-disorder phase
transition when temperature is increased
\cite{Shockley38,Kikuchi74,Binder80}. The ordered phase is the so called
$L1_2$ structure. The fcc lattice can be regarded as four inter-penetrated
simple cubic sublattices (named $\alpha$, $\beta$, $\gamma$ and $\delta$ in
Fig.~\ref{sublatt}); the perfect $L1_2$ order consists in three sublattices
full of $A$-atoms and the other one full of $B$-atoms so that it is fourfold
degenerated. The four equivalent kinds of ordered domains will be called
$\alpha$-, $\beta$-, $\gamma$- and $\delta$-domains according to the
sublattice which contains the minority specie $B$. The order is described by
means of the three following long-range order parameters \cite{Lai90}:

\bea
\nn
\Psi_{1} &= &\frac{2}{N} \sum_{ijk \; \xi \eta \zeta} S_{ijk\; \xi \eta
\zeta}  (-1)^{\xi}
\\
\nn
\Psi_{2} &= &\frac{2}{N} \sum_{ijk\; \xi \eta \zeta} S_{ijk\;\xi \eta
\zeta} (-1)^{\eta}
\\
\Psi_{3} &= &\frac{2}{N} \sum_{ijk\; \xi \eta \zeta} S_{ijk\; \xi \eta
\zeta} (-1)^{\zeta}
\eea
where $S_{ijk\;\xi \eta \zeta}$ is the spin-variable at position
$\vec{r}_{ijk\; \xi \eta \zeta} = a(i + \xi /2, j+ \eta /2, k+\zeta /2)$, and
$i,j,k$ range from $1$ to $L$ and the vector $(\xi , \eta , \zeta)$ take
values $(0,0,0)$, $(0,1,1)$, $(1,0,1)$ and $(1,1,0)$ pointing to the position
of the four sublattices $\alpha$, $\beta$, $\gamma$ and $\delta$
respectively.

\begin{figure}

\centerline{\hbox{
\psfig{figure=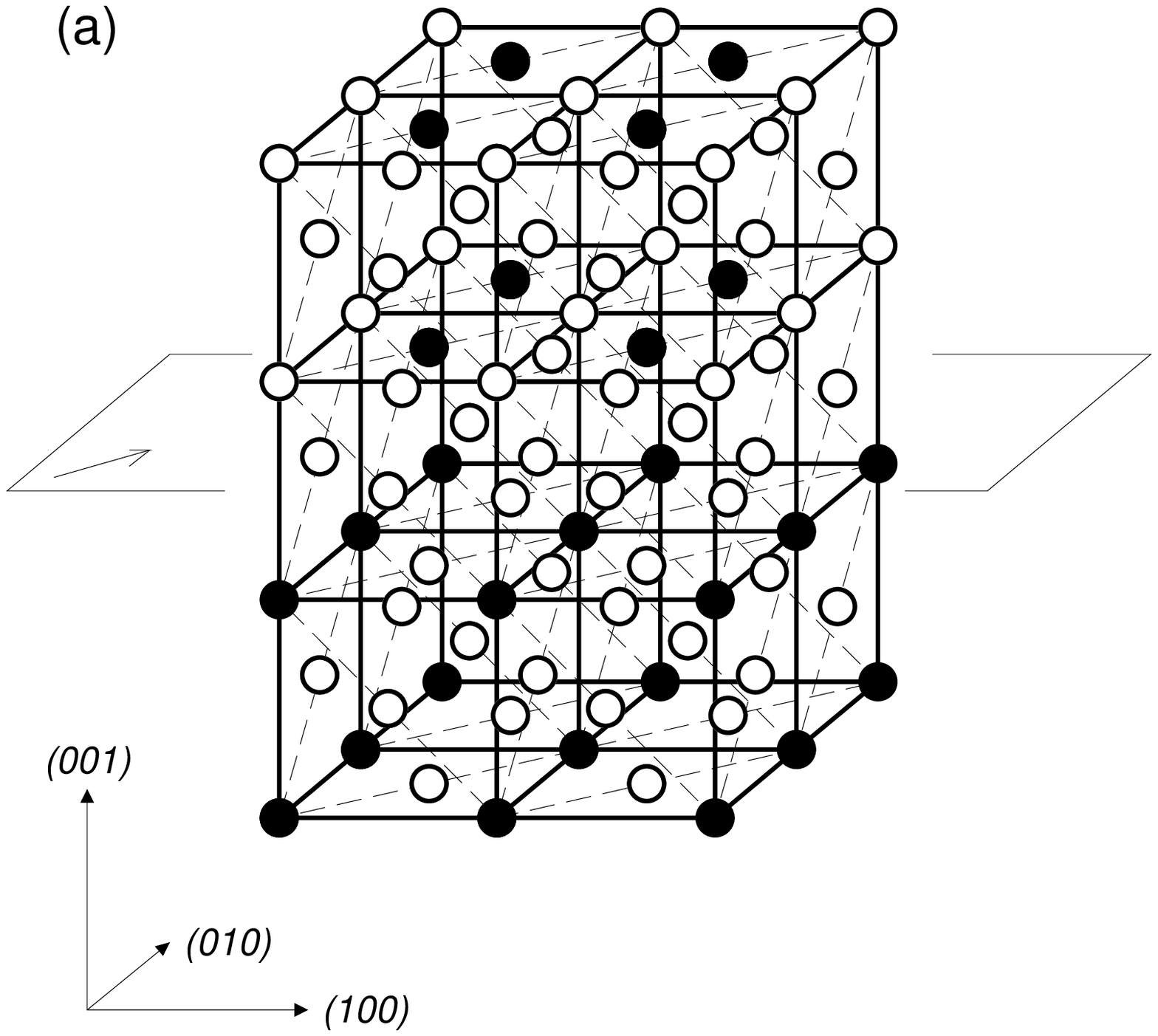,height=276pt,width=300pt}} }

\centerline{\hbox{
\psfig{figure=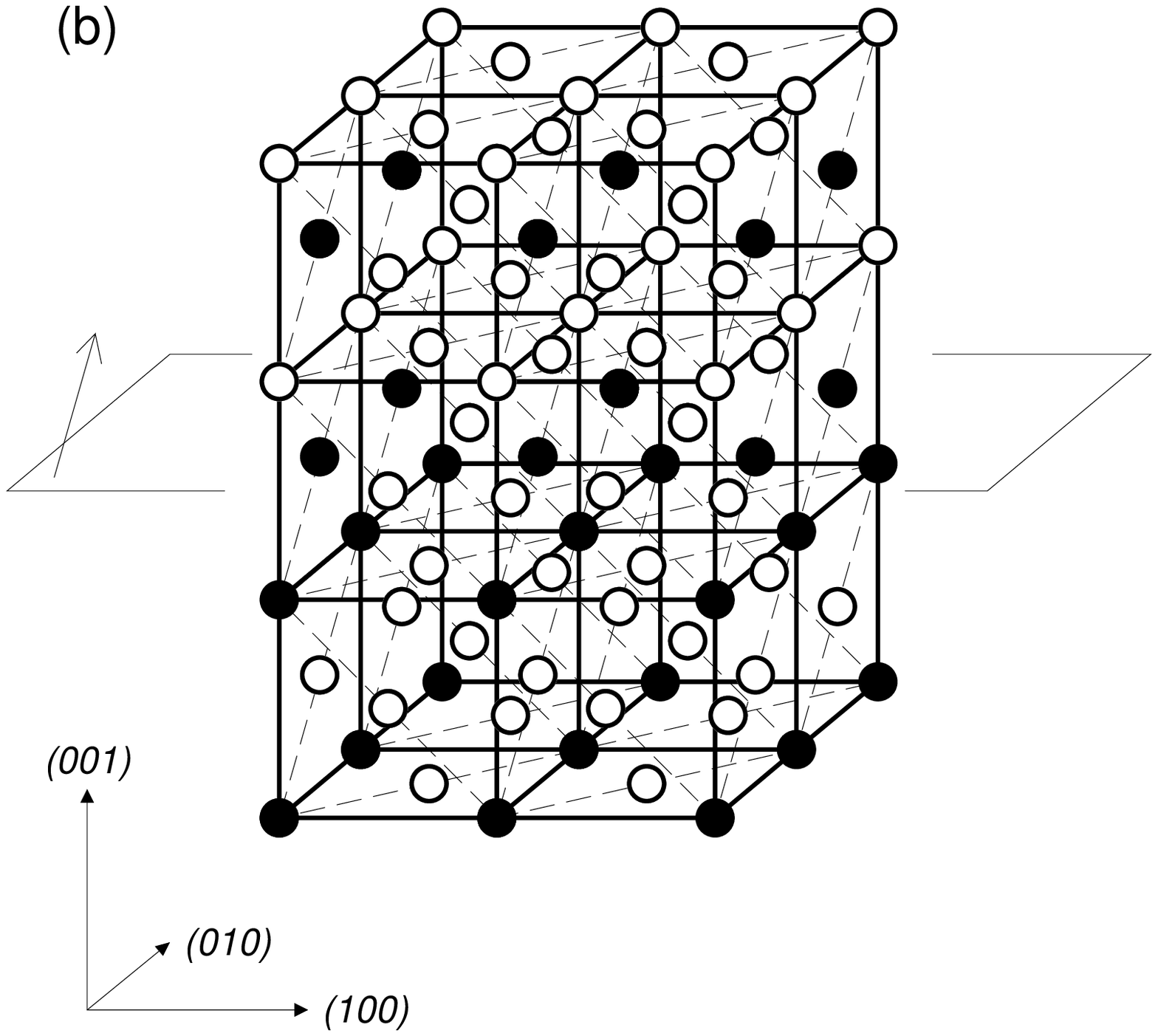,height=276pt,width=300pt}} }

\caption{Schematic picture of the two kinds of APDB explained in the text,
(a) type-1 wall conserving the energy of the system, (b) type-2 wall with
an
excess of unsatisfied bonds. The arrows indicate one of the four equivalents
directions in which the displacement can be made.}

\label{APDB}
\end{figure}

After a quench through the equilibrium transition temperature $T_0$, the four
possible degenerated domains appear and compete during the domain growth
regime. It is well known \cite{Poquette69,Shannon92,Lai90,Warren69,Kikuchi79}
that two kinds of antiphase domain boundaries (APDB) exist (see
Fig.~\ref{APDB}):

\begin{enumerate}

\item The first type (named type-1 or ``half diagonal glide'' walls) of APDB
corresponds to a displacement vector contained in the plane of the APDB. This
kind of boundaries maintain the same number of $A-B$ n.n. bonds as in the
ordered bulk. Hence, in our model with n.n. interactions only, such
boundaries do not suppose any excess of energy. It is also interesting to
remark that they can only appear in specific directions depending on the two
adjacent domains: for instance they can appear perpendicular to direction
[100] between $\alpha$- and $\beta$-domains and between $\gamma$- and
$\delta$-domains. Table \ref{Direc} shows the directions of the type-1 walls
for all the possible neighboring domains. Moreover, it is possible to build
up a structure combining the different kinds of ordered domains with only
type-1 walls, i.e. without excess of energy. At low temperatures, such
structure would not evolve in time.

\item The second type of APDB (type-2 walls) corresponds to a displacement
vector not contained in the boundary plane. Since it does not maintain the
same number of $A-B$ n.n. bonds as in the bulk, it contributes with a
positive excess of energy. It should also be mentioned that these boundaries
contain a local excess of particles (either $A$ or $B$) and they are wider
than the type-1 walls.

\end{enumerate}


\section{Monte Carlo simulation details}
\label{MCDetails}

\subsection{Dynamics}

We have performed two kind of simulation studies:

\begin{enumerate}

\item The first kind includes the simulations of the ordering processes
without vacancies ($N_V = 0$), which have been performed using the standard
Kawasaki dynamics proposing exchanges between n.n. atoms.

\item The second kind includes the simulations with vacancies. In this case
we have used a restricted Kawasaki dynamics proposing n.n. vacancy-atom
exchanges (vacancy jumps) only. This dynamics is more realistic in studying
ordering kinetics in binary alloys. The concentration of vacancies has been
taken the same in all the simulations ($c_V\equiv N_V/N \simeq
3.1\,10^{-5}$).

\end{enumerate}

The concentration of particles is preserved by both dynamics while the order
parameters are not. In both cases we have accepted or refused the proposed
exchange using the usual Metropolis acceptance probability:

\be
p (\Delta {\cal H})=\lt \lbrace \ba{ll} 1 & {\rm if} \; \Delta {\cal H} \le
0 \\
exp\lt \lbrace -\frac{\Delta {\cal H}}{k_BT} \rr \rbrace & {\rm if} \;
\Delta {\cal H} > 0 \ea \rr . ,
\ee
where $\Delta {\cal H}$ is the energy change associated to the proposed
exchange. We define the unit of time, the Monte Carlo step ($mcs$), as the
trial of $N$ exchanges (either atom-atom or vacancy-atom exchanges). In all
the cases we have started the simulations from a completely disordered state,
as would correspond to a system at $T=\infty$. The system is then suddenly
quenched into a final temperature $T_q$ below the transition temperature
$T_0$. To prepare the disordered states we fill up the system with $A$-atoms
and randomly replace $N/4$ of these $A$-atoms by $B$ ones; when it is
necessary, we also choose at random $N_V$ lattice sites to place the
vacancies.

\subsection{Measurements}

Our main interest is the description of the time evolution of the ordered
domains. Usually, the domain size is measured as the inverse of the excess
energy, which is proportional to the amount of interface \cite{Binder85}.
Nevertheless, in our case, there are two coexisting types of interfaces and
only the amount of type-2 walls might be related to the excess energy as has
been discussed in section \ref{MODEL}. Consequently for fcc lattices with
$L1_2$ order it is convenient to simultaneously measure both the excess
energy and the structure factor. These two quantities are defined as follows:

\begin{enumerate}

\item Excess energy per site:

\be
\Delta E(t) \equiv \frac{1}{N} \lt[ {\cal H}(t)-{\cal H}(t \rightarrow
\infty) \rr] ,
\ee
where ${\cal H}(t\rightarrow \infty)$ is the equilibrium energy at the
quenching temperature $T_q$.

\item Structure factor:

\be
S(\vec{k},t) \equiv \lt | \frac{1}{N} \sum_{ijk \; \xi \eta \zeta} S_{ijk
\; \xi \eta \zeta} \exp \lt\{ {\bf i}\frac{2\pi}{a} \vec{k}\cdot\vec{r}_{ijk
\; \xi \eta \zeta} \rr\} \rr |^2,
\ee

where the sum extends over the whole lattice, ${\bf i}$ is the imaginary
constant and $\vec{k}$ is the dimensionless reciprocal vector. The
discreteness of the real space makes the structure factor to be invariant
under the $\langle 200 \rangle$ translations in the reciprocal space. The
periodic boundary conditions imply that the reciprocal space is discrete on a
cubic lattice on the $\vec{k}$-space with lattice spacing ${\ds
\frac{1}{L}}$. Figure \ref{sfac} shows the reciprocal space with the position
of the fundamental and the superstructure peaks.

\end{enumerate}

\subsection{Domains and superstructure peak}

The temporal evolution of the domain structure is illustrated in
Fig.~\ref{plans}(a,b,c). Three snapshots, at selected times, corresponding to
a section parallel to the (100) planes are presented. The four possible
ordered regions are indicated with different colors. Flat (type-1) and curved
(type-2) interfaces can be observed. The mean distance between such walls is
related to the shape of the superstructure peaks of the structure factor.
Discrepancies in the shape of these peaks have been reported in the
literature. Old x-ray measurements by Cowley \cite{Cowley50} suggested that
the peaks are square shaped. More recently it has been insinuated that the
peaks are disk shaped \cite{Shannon92}. Our simulations give square (or even
star-like) shaped peaks as it can be seen in the temporal sequence presented
in Fig.~\ref{plans}(d,e,f). Such anisotropy in the peak shape does not
necessarily imply that the ordered domains are needle shaped, but arises from
the correlation between the ordered domains. For instance the superstructure
peak at (100) position accounts for the long range order parameter
$\Psi_{1}$. This order parameter does not discriminates between $\alpha$- and
$\beta$-domains (both having $\Psi_1=-1$) nor between $\gamma$- and
$\delta$-domains (both having $\Psi_1=1$). The tendency of the domains to
locate in such a way that the interfaces does not have extra energy favors
the formation of $\alpha$-$\beta$ and $\gamma$-$\delta$ boundaries (type-1)
perpendicular to the [100] direction. Therefore the regions with a high value
of the order parameter $\Psi_{1}$ are anisotropic, producing the anisotropy
of the $(100)$ peak. The same happens for the other peaks at $(010)$ and
$(001)$.

\begin{figure}

\centerline{\hbox{
\psfig{figure=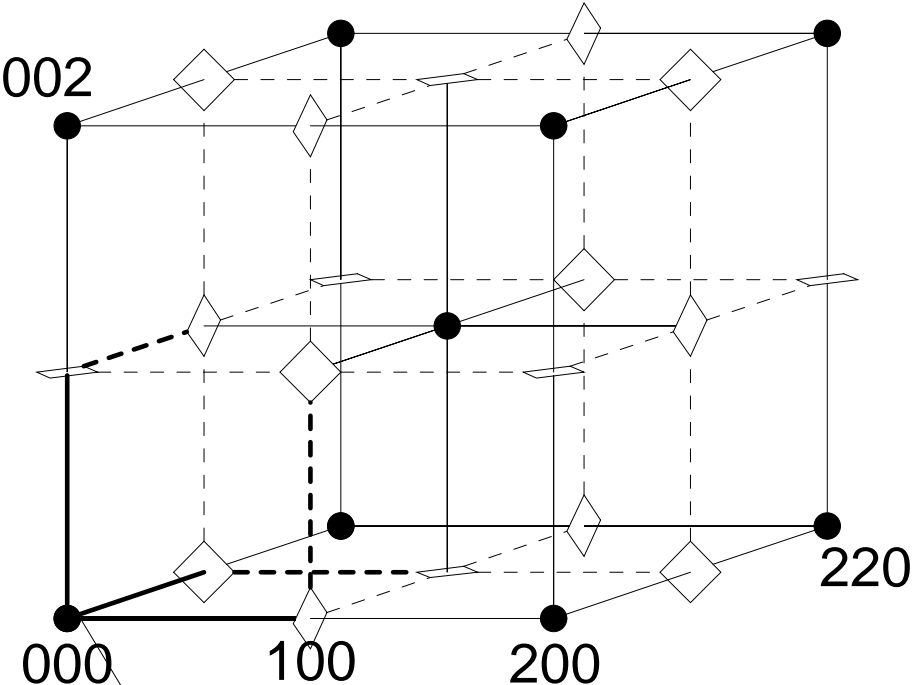,height=300pt,width=400pt}} }

\caption{Reciprocal $\vec{k}$-space showing the fundamental (filled circles)
and the superstructure (squares) peaks. The average of the structure factor
on the thick solid lines is $S_r(q,t)$ and the average on the thick dashed
lines is $S_t(q,t)$.}

\label{sfac}
\end{figure}

\begin{figure}

\centerline{\hbox{
\psfig{figure=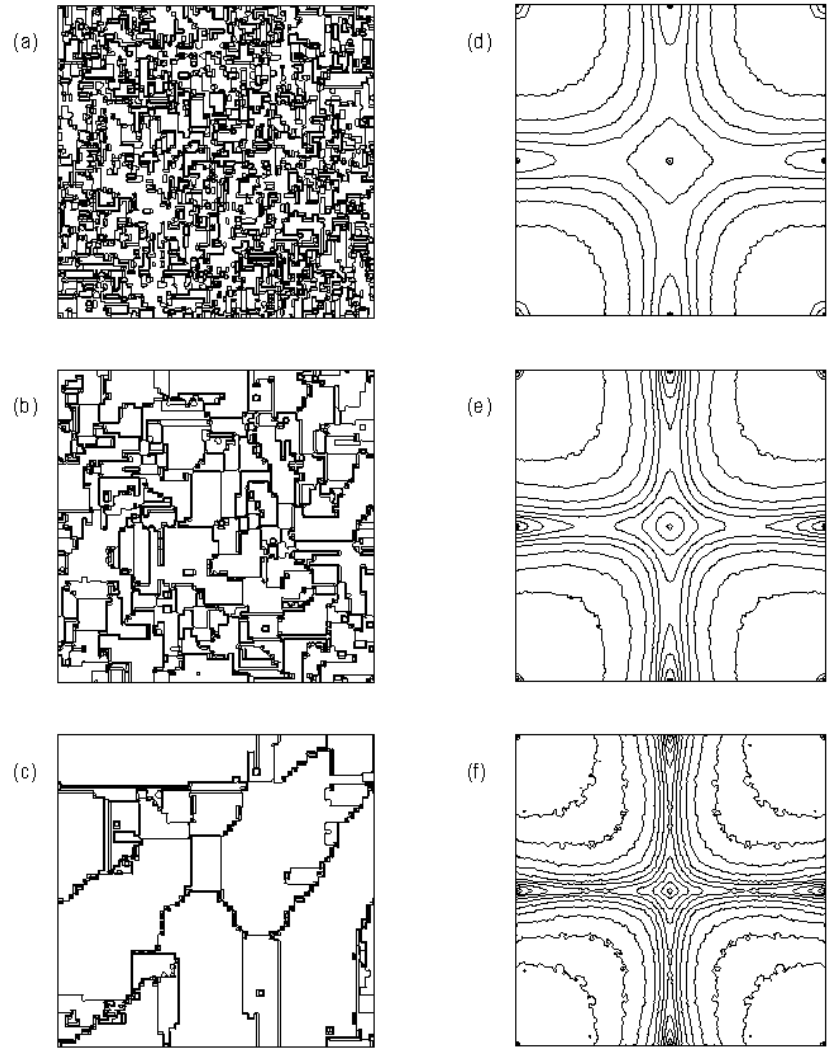,height=418pt,width=314pt}} }

\caption{Snapshots of the system evolution in real space [(a), (b) and (c)]
and in reciprocal space [(d), (e) and (f)]. The real space pictures,
corresponding to a section perpendicular to (100) direction, show in four
different colors the structure of ordered domains. The reciprocal space plots
show the structure factor maps in the plane (100) with color scale increasing
logarithmically from violet to yellow. Data corresponds to a simulation using
the atom-atom exchange mechanism at $T_q=1.0$, system size $L=64$, and times
$t=18\;mcs$ [(a) and (d)], $t=198\;mcs$ [(b) and (e)] and $t= 1998\;mcs$ [(c)
and (f)].}

\label{plans}
\end{figure}

The inverse of the amplitudes of the superstructure peak along both, the
radial ($\sigma_r$) and the transverse ($\sigma_t$) directions, are related
to the mean distance between type-2 and type-1 walls respectively. The
measurement of $\sigma_r$ has been done by analyzing the profile of:

\bea
\nn
S_r(q,t)& = & \frac{1}{3}\lt [S\lt((1,0,0)-\frac{1}{L}
(q,0,0),t \rr)+ \rr .
S \lt( (0,1,0)-\frac{1}{L} (0,q,0),t\rr)+ \\
& & \lt. S\lt( (0,0,1)-\frac{1}{L} (0,0,q),t\rr) \rr] ,
\eea
where $q=0,1,\dots,L$ is the distance from the superstructure peak (in units
of ${\ds \frac{1}{L}}$). Following the notation in Ref.~\cite{Shannon92}, we
will refer to this profile as the radial scan of the structure factor. It
corresponds to the average on the three thick continuous lines of
Fig.~\ref{sfac}. The measurement of $\sigma_t$ has been done by analyzing the
profile of:

\bea
\nn
S_t(q,t) & = & \frac{1}{6}\lt [ S\lt( (1,0,0)+\frac{1}{L}
(0,0,q),t \rr)+
S\lt( (1,0,1)-\frac{1}{L} (0,0,q),t \rr)+ \rr.\\
\nn
& & S\lt( (0,1,0)+\frac{1}{L} (q,0,0),t \rr)+
S\lt( (1,1,0)-\frac{1}{L} (q,0,0),t \rr)+ \\
& & \lt. S\lt(  (0,0,1)+\frac{1}{L} (0,q,0),t \rr)+
S\lt( (0,1,1)-\frac{1}{L} (0,q,0),t \rr) \rr] ,
\eea
where $q=0,1,\dots,L/2$. We will refer to it as transverse scan of the
structure factor. It corresponds to the average on the two symmetric parts of
the three thick dashed lines of Fig.~\ref{sfac}. It can be easily shown that
due to the symmetries of the structure factor the average over the continuous
(dashed) thick lines is equal to the average over all the continuous (dashed)
lines of the Fig.~\ref{sfac}. In addition, at every time both profiles have
been averaged over a certain number of independent runs (about 25 for $L=64$
and 40 for the other system sizes). This last average is indicated by
means of angular brackets ($\langle \cdots \rangle $). The finite size
effects have been studied by simulating systems of linear sizes $L=$ 20, 28,
36 and 64 ($N= $ 32000, 87808, 186624 and 1048576 sites respectively). We
have also studied the effect of the quenching temperature performing
simulations at $T_q=1.0J/k_B$ ($T_q/T_0\simeq 0.55$) and $T_q= 1.5J/k_B$
($T_q/T_0 \simeq 0.83$).

\subsection{Fitting procedure for $\sigma_r$ and $\sigma_t$}

We have measured $\sigma_r$ and $\sigma_t$ using the two following methods.

\begin{enumerate}

\item The first method is based on the evaluation of the second moment of the
scan:

\be
\sigma^2(t) \equiv \frac{\ds \sum_{q=0}^{q_{max}} q^2\langle S(q,t)
\rangle}{\ds \sum_{q=0}^{q_{max}}\langle S(q,t)\rangle}
\label{sigma}
\ee
where $S$ represents either $S_r$ or $S_t$, and $q_{max}$ is the first
$q$-value for which $\langle S(q,t) \rangle$ is lower  than a background
threshold. The value of this background has been taken double of the mean
value of the structure factor of a completely disordered system (excluding
the fundamental peak).

\label{met1}

\item The second method is based on the fitting of a lorentzian function to
the data of the corresponding scan. In order to account for the large-$q$
tail of the structure factor we have fitted $\log \langle S(q,t) \rangle $ to
the three-parameters ($a$, $\sigma$, $B$) function:

\be
\log \langle S(q,t) \rangle \simeq \log \lt \{ \frac{a}{\lt [
1+{\ds \lt( \frac{q}{\sigma(t)} \rr )^2} \rr ]^\alpha } +B \rr \},
\label{fitt}
\ee
where $\sigma$ is the estimation of the width, $B$ is the background, $a$ is
the fitted intensity and $\alpha$ is an exponent (not fitted) that we discuss
in the next paragraph.

\label{met2}

\end{enumerate}

We have analyzed the validity of both methods for the two scans of the
structure factor. The first method turns out to be adequate only for the
radial scans. This is because the large-$q$ tail of the transverse scan
decays very slowly and, therefore, $\sigma_t$ is strongly affected by the
choice of $q_{max}$. The second method can be used for both scans. Moreover,
both methods render equivalent results for radial scans. Therefore we will
estimate $\sigma_r$ using method \ref{met1} and $\sigma_t$ using method
\ref{met2}. Concerning the exponent $\alpha$ we have tried $\alpha= 1, 2$ and
$3/2$ for both scans. In general the best fits to the radial scan have been
obtained with $\alpha= 3/2$, while for the transverse scans $\alpha=1$ gives
the best results.

\subsection{Scaling}

We have tested the existence of dynamical scaling in both the radial and the
transverse directions by plotting the corresponding
scaling function $\tilde{S}(\tilde{q})$ defined from the following
expression:


\be
\langle S(q,t) \rangle = \frac{1}{\sigma(t)^d}
\tilde{S} (\tilde{q}) ,
\label{escr}
\ee
where $\displaystyle \tilde{q}=\frac{q}{\sigma(t)}$ is the scaling variable
and $d$ is the space dimensionality. Scaling has been tested by checking the
overlap of the data corresponding
to different times and also to different system sizes. Theoretical
predictions for the scaling function $\tilde S (\tilde q)$ exist
\cite{Mazenko91,Lai90}, specially concerning the behaviour for large values
of $q$.


\section{Results}
\label{Results}

For the sake of clarity the results are presented in the following order:
first, the results corresponding to equilibrium simulations (subsection A);
second, those concerning the evolution of the system using the atom-atom
exchange mechanism (subsection B); and third, the results obtained by means
of the vacancy-atom exchange mechanism (subsection C). The results in
subsections B and C are always given at quenching temperatures $T_q=
1.0J/k_B$ and $T_q=1.5J/k_B$. The structure factor profiles are only
presented for $L=64$ although data corresponding to smaller $L$ have also
been analyzed.

\subsection{Equilibrium}

Starting from perfectly ordered systems with $L=20$, we have step-by-step
heated them from $T=0.5J/k_B$ to $T= 3.0J/k_B$ and cooled them again down
to $T=0.5J/k_B$. At each temperature we have let the system to reach
equilibrium (after $\sim 12 \; 10^3 mcs $) and have obtained the equilibrium
energy ${\cal H}(t\rightarrow \infty)$ and the mean long range order
parameter defined as:

\be
\Psi \equiv \frac{|\Psi_1|+ |\Psi_2|+|\Psi_3|}{3} .
\ee

\begin{figure}

\centerline{\hbox{
\psfig{figure=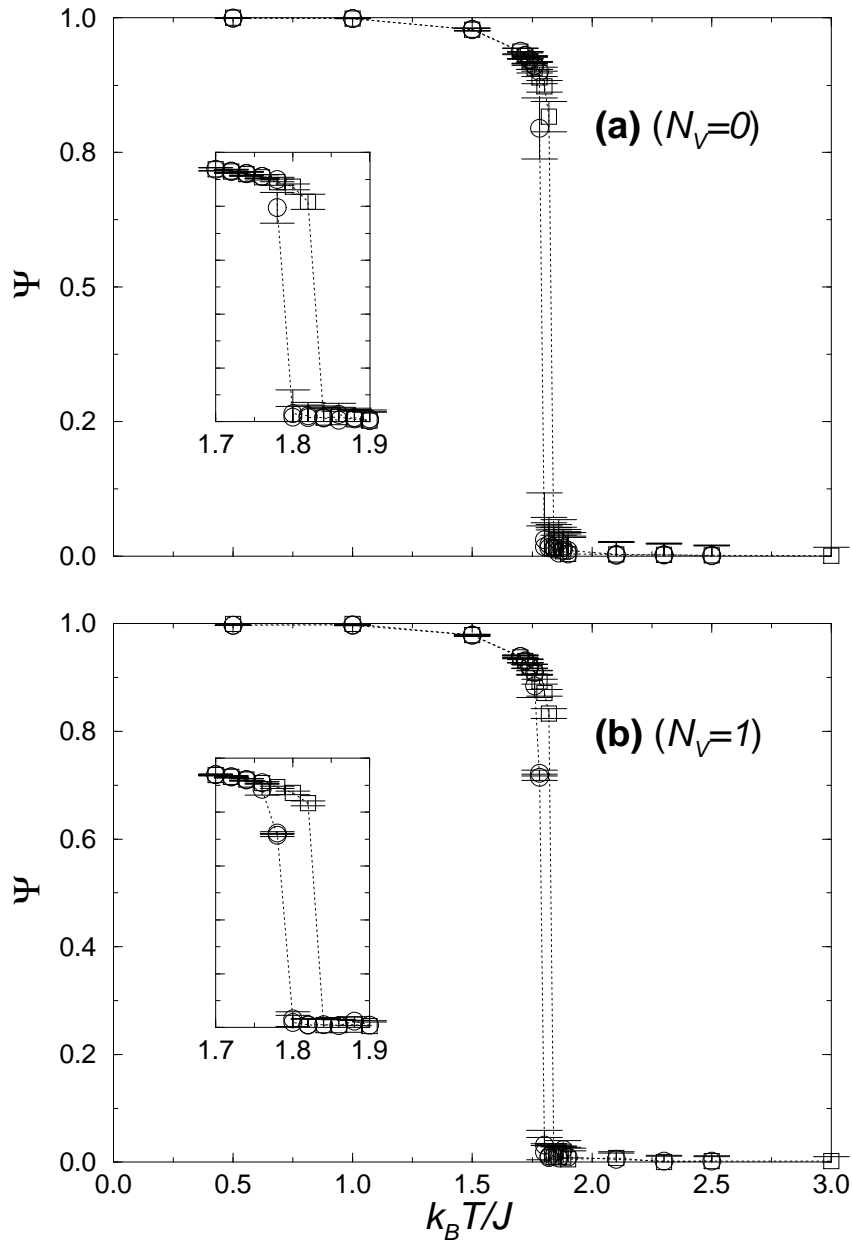,height=513pt,width=351pt}} }

\caption{Equilibrium order parameter as a function of temperature for $L=20$
with (a) $N_V=0$ (atom-atom exchanges) and (b) $N_V=1$ (vacancy-atom
exchanges). The insets show in detail the hysteresis due to the discontinuous
character of the transition. From both figures it can be concluded that $k_B
T_c / J=1.81 \pm 0.03$. Squares correspond to a heating process while circles
to a cooling one.}

\label{equil}
\end{figure}

The temporal average of this order parameter is presented in Fig.~\ref{equil}
as a function of temperature. No differences can be observed between the
cases corresponding to $N_V= 0$ (with the atom-atom exchange mechanism) and
$N_V=1$ (with the vacancy-atom exchange mechanism). We have found that the
transition is first order and that the heating-cooling cycle shows
hysteresis. The transition temperature has been estimated to be $T_0=1.81 \pm
0.03  J/k_B$ compatible with the Monte Carlo results given in
Ref.~\cite{Binder80}.

\subsection{Atom-atom exchange mechanism}

\begin{figure}

\centerline{\hbox{
\psfig{figure=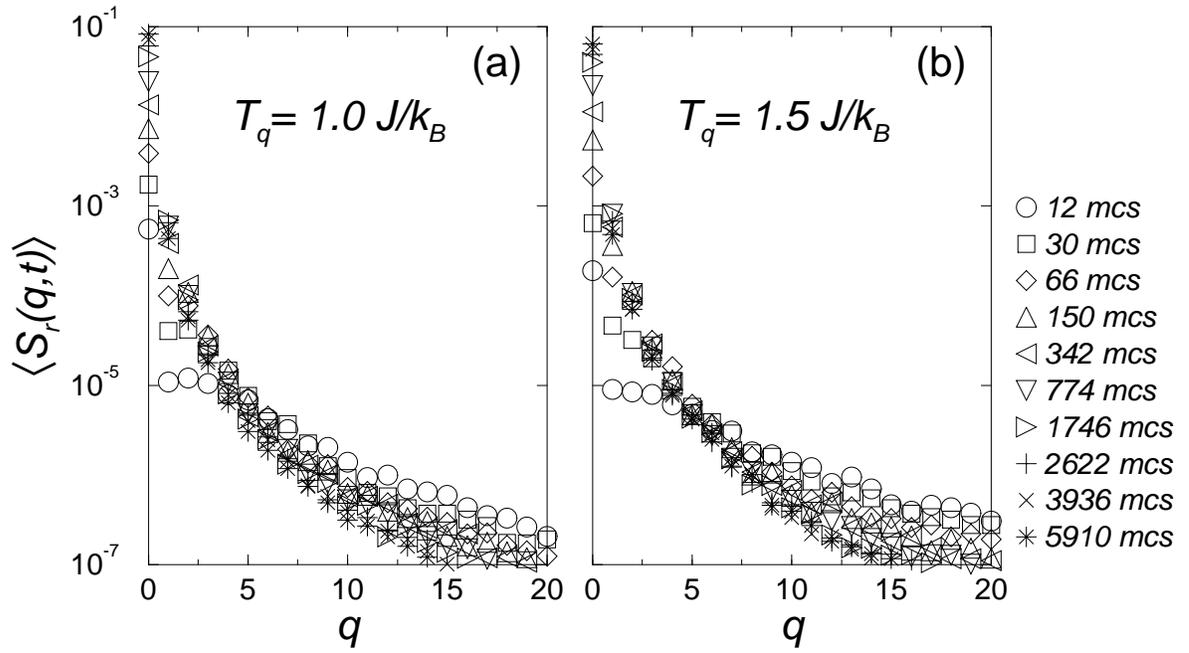,height=305pt,width=400pt}} }

\caption{Linear-log plot of the averaged radial scan of the structure
factor at different times and quenching temperatures $T_q= 1.0 J/k_B$ (a)
and $T_q=1.5J/k_B$ (b). Both graphs correspond to systems of linear size
$L=64$ for the atom-atom exchange mechanism.}

\label{facr.sec}
\end{figure}

\begin{figure}

\centerline{\hbox{
\psfig{figure=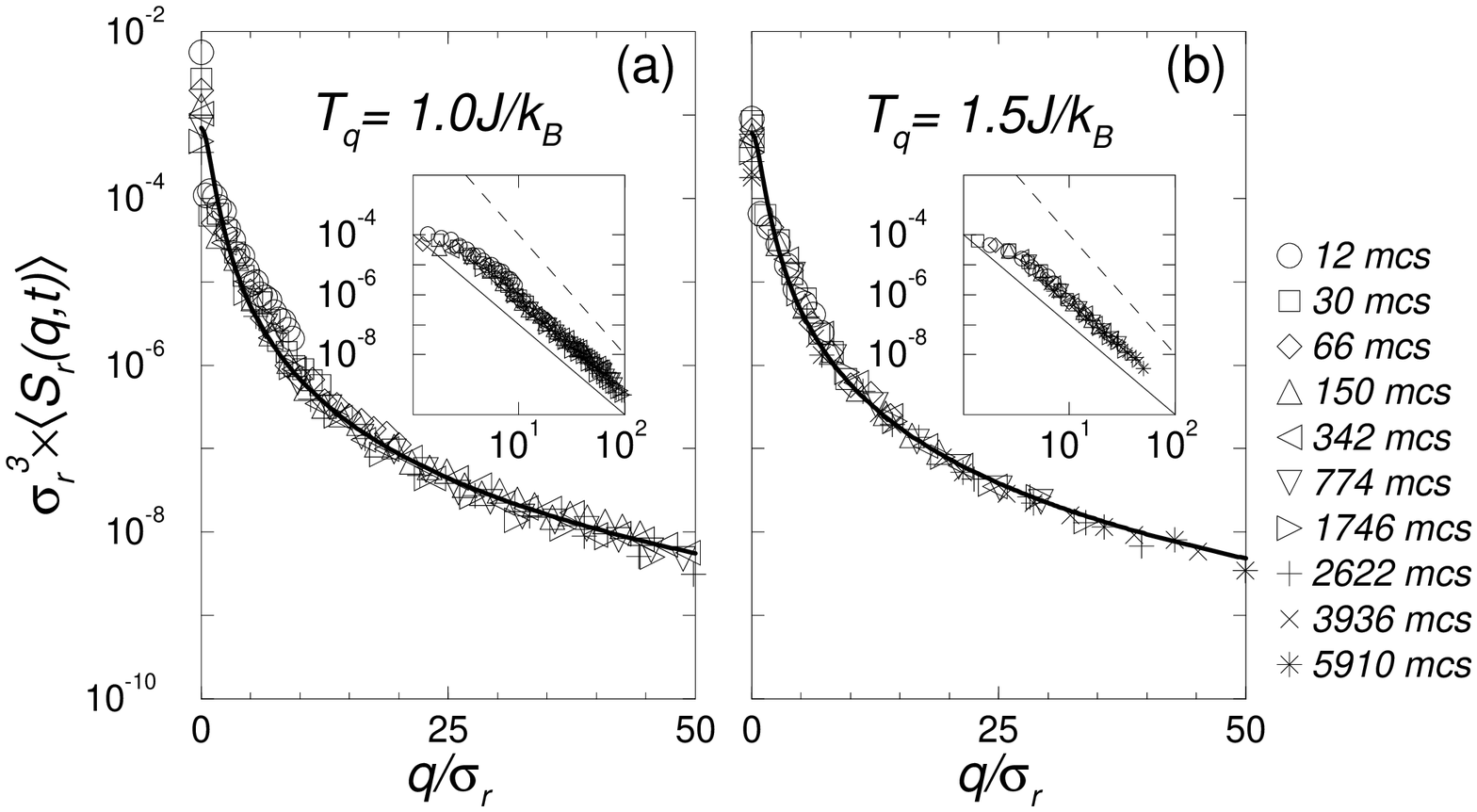,height=305pt,width=400pt}} }

\caption{Linear-log and log-log (insets) plots of the scaled radial scan of
the structure factor at different times and quenching temperatures $T_q= 1.0
J/k_B$ (a) and $T_q= 1.5 J/k_B$ (b). Data corresponds to systems of linear
size $L=64$ for the atom-atom exchange mechanism. The solid thick line
corresponds to a fit of expression (3.7) with $\alpha=1.5$ and $\sigma=1$.
The dashed lines in the insets show the Porod's law
$\tilde{S}_r(\tilde{q})\sim \tilde{q}^{-4}$, while solid lines show the slope
of $\tilde{q}^{-3}$.}

\label{facescr.sec}
\end{figure}

Figures \ref{facr.sec}(a) and  \ref{facr.sec}(b) show the profiles, at
different times, of the radial scan of the structure factor for the two
studied quenching temperatures. The corresponding structure factors, scaled
according to eq. (\ref{escr}), are shown in Fig.~\ref{facescr.sec}. In
general, the overlap of the different curves is quite good. It has been
checked that data corresponding to different system sizes also fall on the
same curve. Nevertheless, deviations from this scaling can be well
appreciated at $q=0$ in Fig.~\ref{facescr.sec}(a) ($T_q=1.0J/k_B$). We will
come back to this point in the discussion. The continuous lines show fits of
lorentzian functions [eq.~(\ref{fitt})] with $\alpha=3/2$ and $\sigma=1$
which corroborates the validity of such kind of fitting function for all the
individual profiles. The insets in Fig.~\ref{facescr.sec}(a) and
Fig.~\ref{facescr.sec}(b) display log-log plots of the scaled radial scans:
for large $\tilde{q}$-values $\tilde{S}_r(\tilde{q})$ decays as
$\tilde{q}^{-3}$, as indicated by a continuous line.

\begin{figure}

\centerline{\hbox{
\psfig{figure=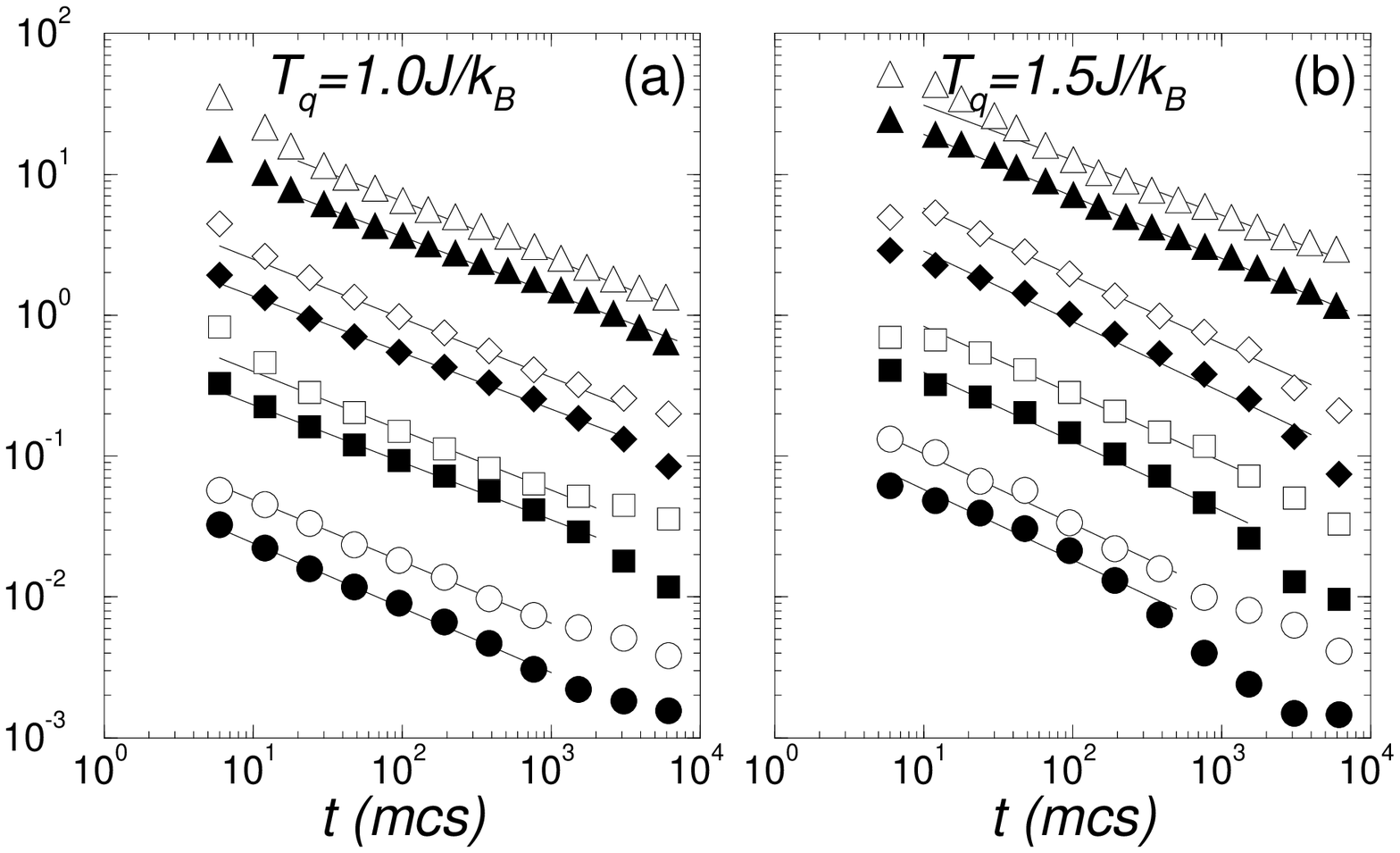,height=305pt,width=400pt}} }

\caption{Time evolution, in the case of atom-atom exchange mechanism, of
$\langle\Delta E(t)\rangle$ (filled symbols) and of $\sigma_r(t)$ (open
symbols), for $T_q= 1.0J/k_B$ (a) and $T_q=1.5J/k_B$ (b) for systems of
$L=20$ (circles), $L=28$ (squares), $L=36$ (diamonds) and $L=64$ (triangles).
The solid lines are the best power-law fits. Data corresponding to different
sizes have been vertically shifted to clarify the picture.}

\label{ener.sec}

\end{figure}

Figures \ref{ener.sec}(a) and \ref{ener.sec}(b) show  log-log plots of the
time evolution of $\langle \Delta E(t)\rangle$ and $\sigma_r(t)$ for
different system sizes and the two studied quenching temperatures. Solid
straight lines are the following fitted power-laws:

\bea
\nn
\langle \Delta E(t) \rangle & \sim & t^{-x} \\
\sigma_r(t) & \sim & t^{-y} .
\label{plr}
\eea

\begin{figure}

\centerline{\hbox{
\psfig{figure=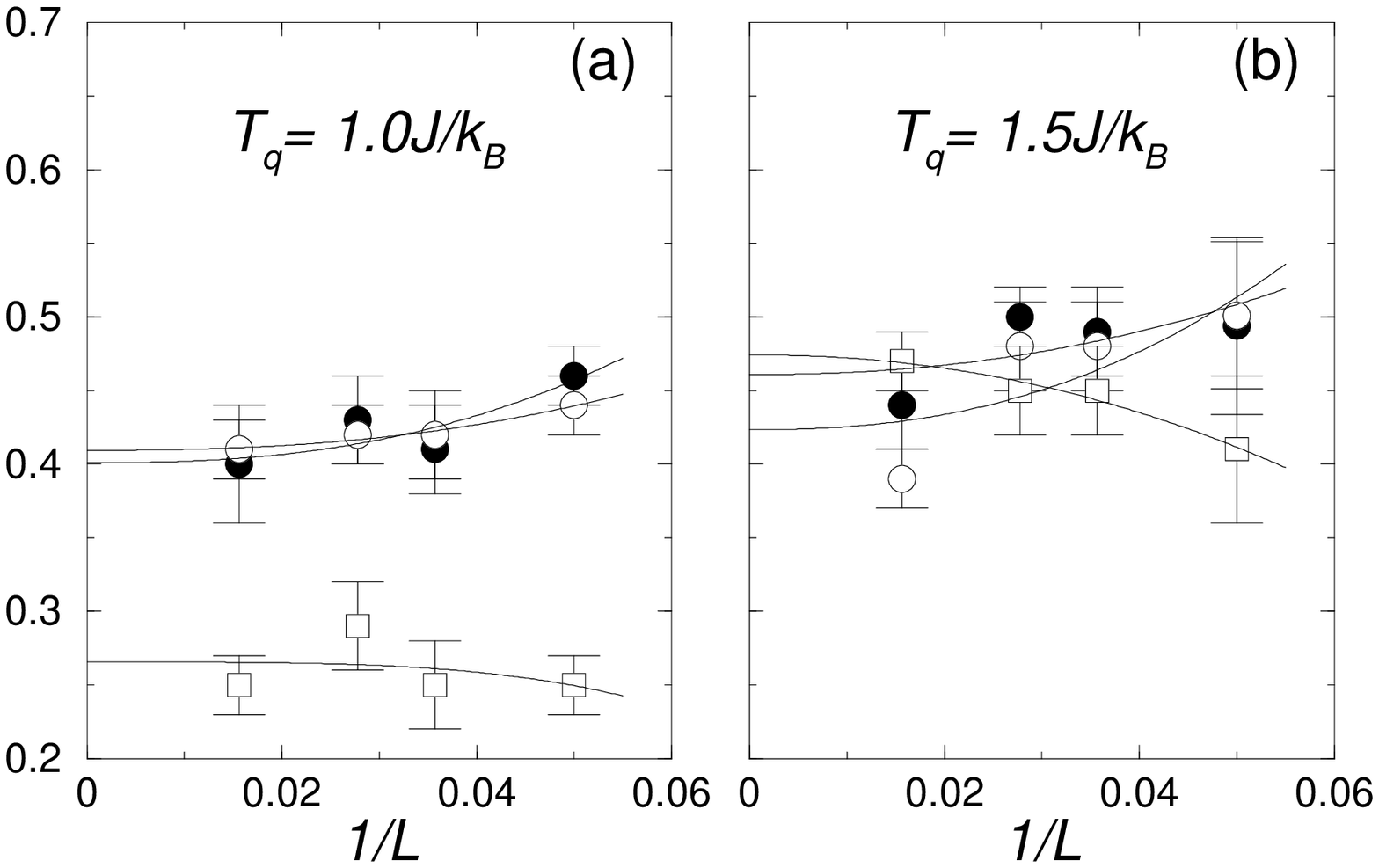,height=305pt,width=400pt}} }

\caption{Exponents $x$ ($\bullet$) from the excess energy per site ($\Delta
E$), $y$ ($\circ$) from the second moment of the radial scan of the
superstructure peak $\sigma_r$ and $z$ ($\Box$) from the fitted amplitude of
the transverse scan of the superstructure peak $\sigma_t$, as a function of
$1/L$ with the atom-atom exchange mechanism. Data corresponds to temperature
$T_q=1.0J/k_B$ (a) and $T_q=1.5J/k_B$ (b). Solid lines show the best fits of
eq. (5.2).}

\label{exp.sec}
\end{figure}

The behaviour of the growth-exponents $x$ and $y$ for the two quenching
temperatures are presented in Fig.~\ref{exp.sec} in front of $1/L$. The
corresponding numerical values are listed in Table \ref{tabsec}. The
estimations of the two growth exponents, $x$ and $y$, are coincident within
the errors bars. A general tendency of the exponent to increase when
temperature is increased and to decrease when increasing the system size is
observed. Extrapolation to $L\rightarrow \infty$, following the method
explained in section \ref{Discussion}, renders $x\simeq y\simeq 0.40$ for
$T_q =1.0J/k_B$ and $x\simeq y\simeq 0.44$ for $T_q =1.5J/k_B$

\begin{figure}

\centerline{\hbox{
\psfig{figure=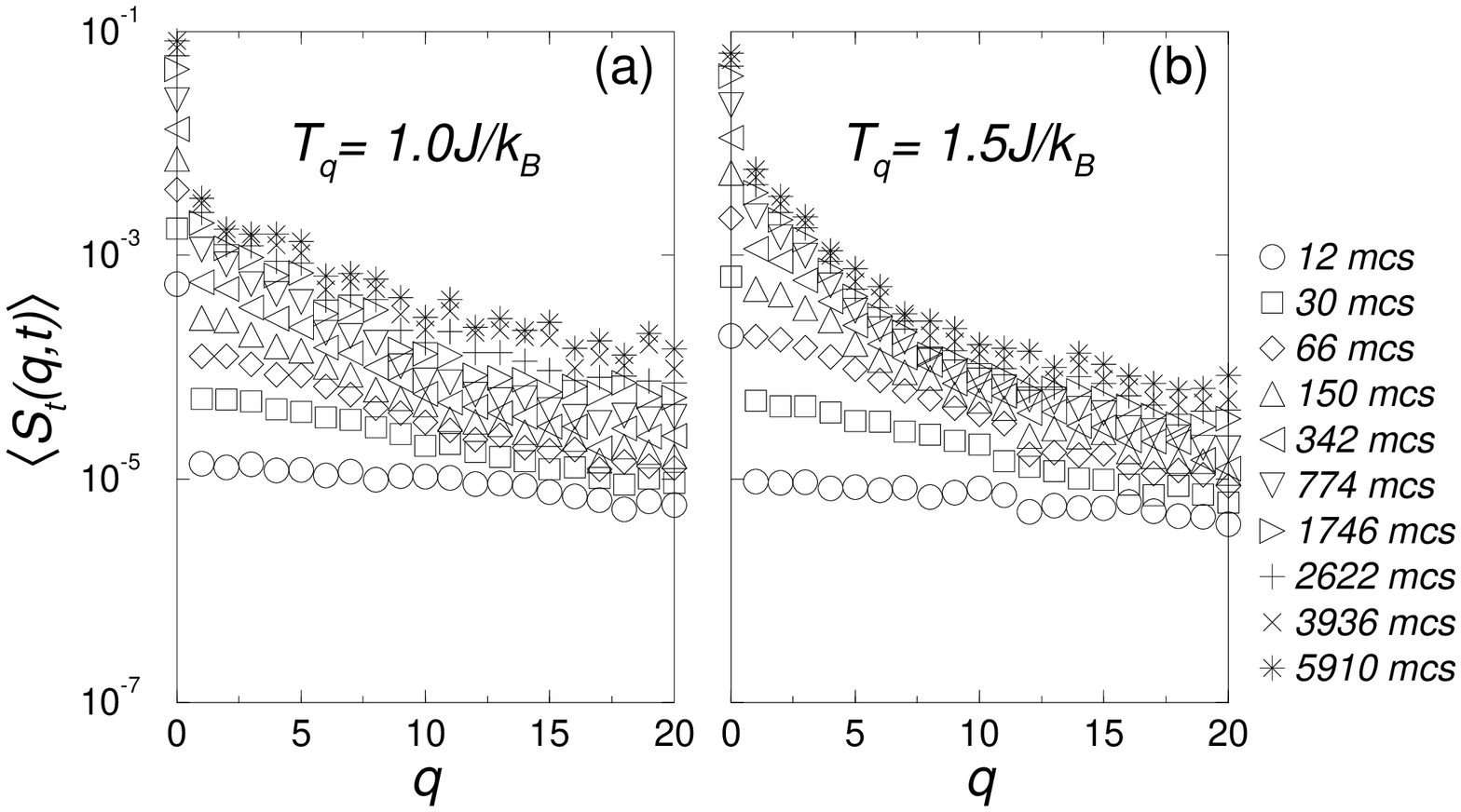,height=305pt,width=400pt}} }

\caption{Linear-log plot of the averaged transverse scan of the structure
factor at different times and quenching temperatures $T_q= 1.0 J/k_B$ (a)
and $T_q=1.5J/k_B$ (b) with the atom-atom exchange mechanism. Data
correspond to systems of linear size $L=64$.}

\label{fact.sec}
\end{figure}

\begin{figure}

\centerline{\hbox{
\psfig{figure=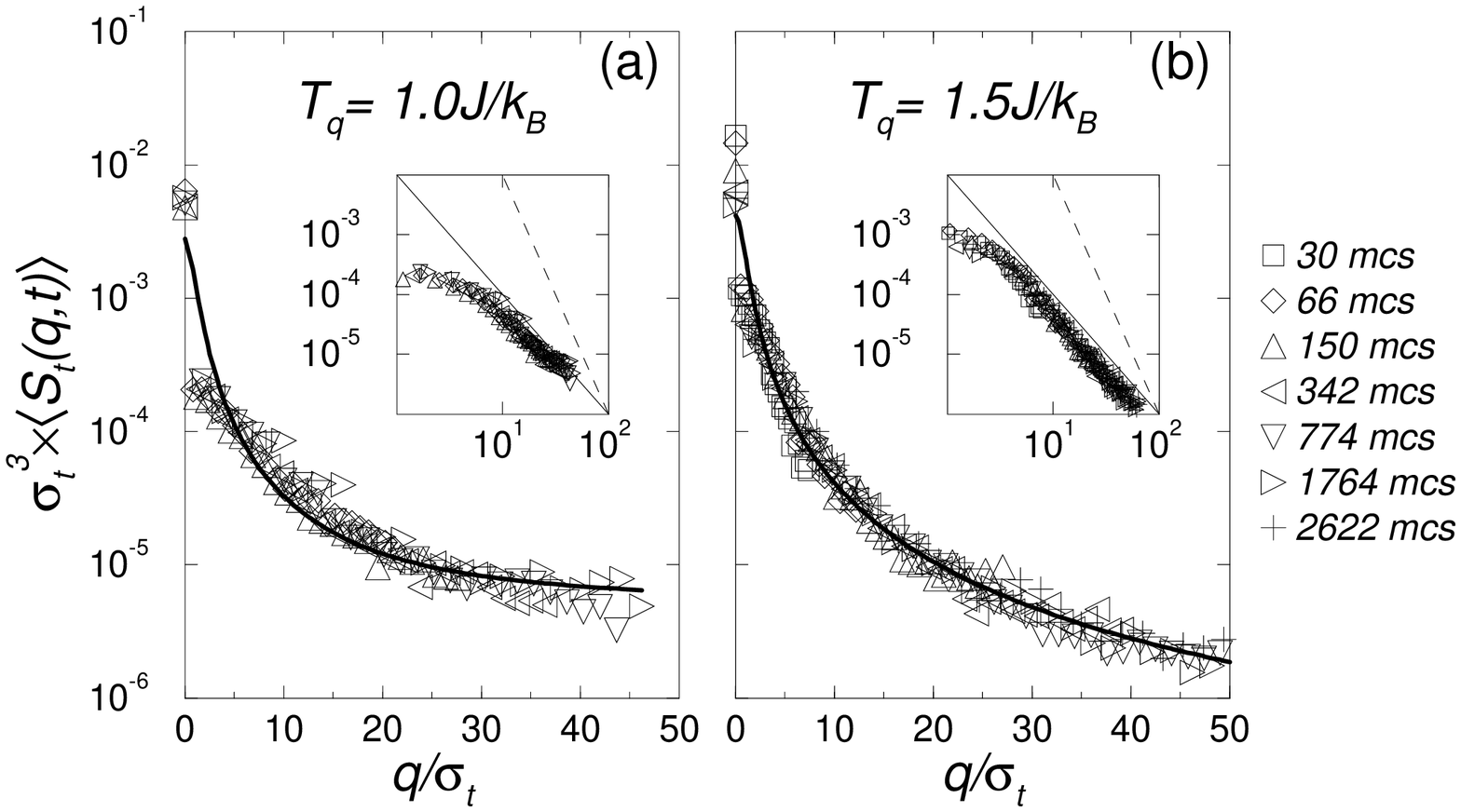,height=305pt,width=400pt}} }

\caption{Linear-log and log-log (insets) plots of the scaled transverse scan
of the structure factor at different times and quenching temperatures $T_q=
1.0 J/k_B$ (a) and $T_q=1.5J/k_B$ (b) with atom-atom exchange mechanism. Data
correspond to systems of linear size $L=64$. The solid thick lines correspond
to fits of expression (3.7) with $\alpha=1$ and $\sigma=1$. The dashed lines
in the insets show the Porod's law $\tilde{S}(\tilde{q})\sim \tilde{q}^{-4}$,
while the solid lines show the slope of $\tilde{q}^{-2}$.}

\label{facesct.sec}
\end{figure}

Figure \ref{fact.sec} shows linear-log plots of the transverse scan of the
structure factor for the two studied quenching temperatures, at different
times. The corresponding scaled transverse scans are shown in
Fig.~\ref{facesct.sec}. The overlap of the curves is rather satisfactory,
however it is not as broaden in time as for the radial scan case. Note that
in Fig.~\ref{facesct.sec}(b) lack of scaling at $q=0$ is clearly evident.
As for the radial scan case, we have also verified the overlap of the data
corresponding to systems with different sizes. The continuous lines show fits
of  lorentzian functions [eq.~(\ref{fitt})] with $\alpha=1$ and $\sigma=1$.
The insets display log-log plots of these scaled transverse scans: in this
case $\tilde{S}_t(\tilde{q})$ decays as $\tilde{q}^{-2}$ for large
$\tilde{q}$, as indicated by the continuous line.

\begin{figure}

\centerline{\hbox{
\psfig{figure=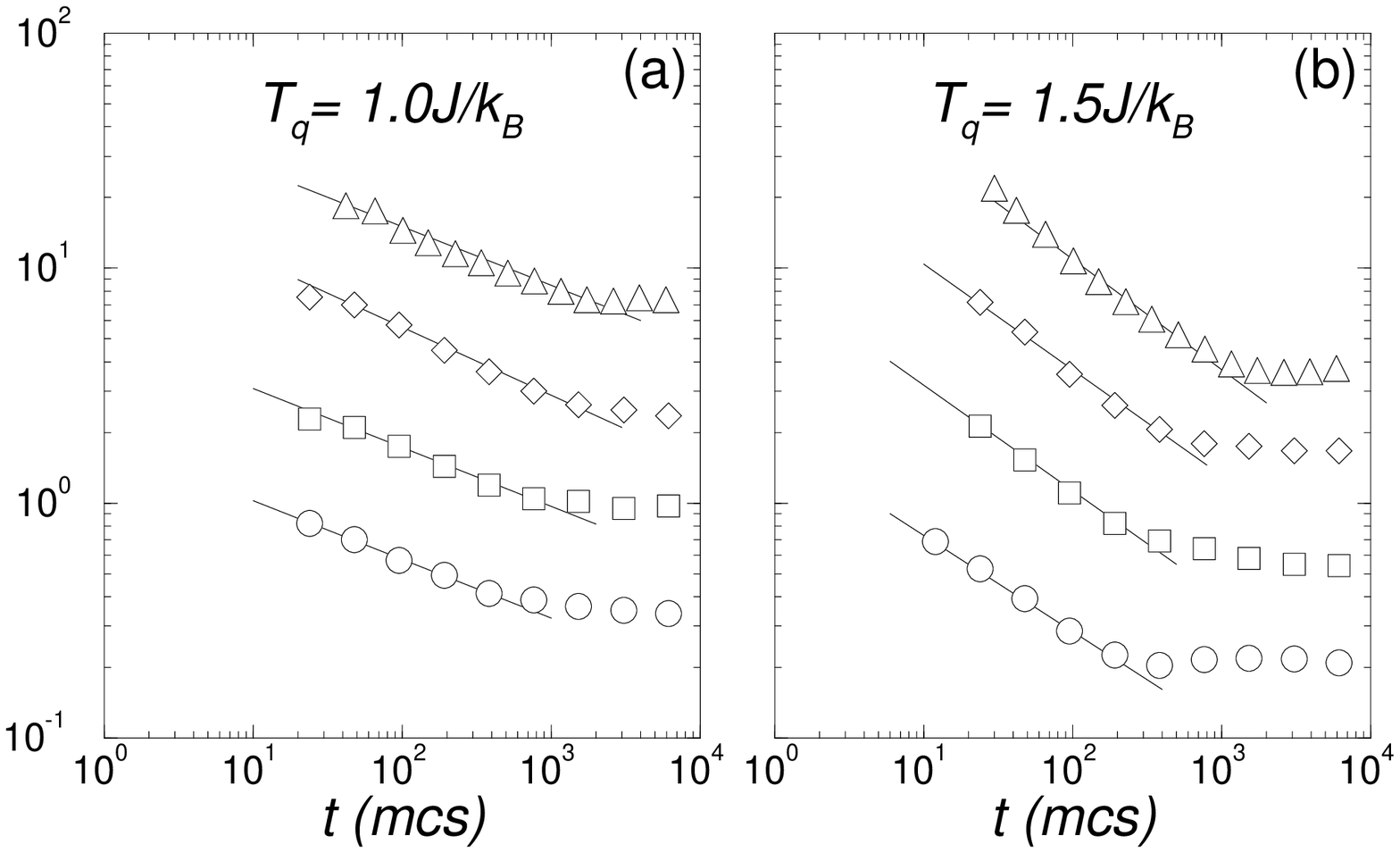,height=305pt,width=400pt}} }

\caption{Log-log plot of the mean distance between type-1 walls for $L=20$
($\circ $), $L=28$ ($\Box$), $L=36$ ($\Diamond $), $L=64$ ($\triangle$) with
atom-atom exchange mechanism. Data correspond to quenching temperatures $T_q=
1.0J/k_B$ (a) and $T_q=1.5J/k_B$ (b). Solid lines show the best power-law
fits.}

\label{elengtht.sec}

\end{figure}

Log-log plots of the time evolution of $\sigma_t(t)$ for different system
sizes and the two studied quenching temperatures are presented in
Fig.~\ref{elengtht.sec}. Solid straight lines correspond to the fitted
power-law:

\be
\sigma_t(t)\sim t^{-z}
\label{plt}
\ee
The values of the growth exponents are listed in Table \ref{tabsec} and
plotted in front of $1/L$ in Fig.~\ref{exp.sec}. In this case, the
extrapolations to $L\rightarrow \infty$ render $z \simeq 0.26$ for $T_q
=1.0J/k_B$ and $z \simeq 0.47$ for $T_q=1.5J/k_B$. The time evolution of the
order parameter, i.e. the structure factor at $q=0$ ($\langle {\displaystyle
S_r(q=0,t) \rangle = \langle S_t(q=0,t) \rangle = \left \langle
\frac{\Psi_1^2+\Psi_2^2+\Psi_3^2}{3} \right \rangle } $), is shown in
Fig.~\ref{smax.sec}.

\subsection{Vacancy-atom exchange mechanism}

The profiles of the radial scans of the structure factor can be seen in
Fig.~\ref{facr.vac}. The same data scaled according eq.~(\ref{escr}) are
shown in Fig.~\ref{facescr.vac}. The overlap of the different curves
corroborates the scaling hypothesis. It is worth noting that scaling holds
even at $q=0$, contrarily to the case of atom-atom exchange mechanism at low
quenching temperature. Scaling of data corresponding to different system
sizes has also been checked. The continuous lines show fits of lorentzian
functions [eq.~(\ref{fitt})] with $\alpha=3/2$ and $\sigma=1$. The log-log
plots shown in the insets of Fig.~\ref{facescr.vac} reveal that, for large
$\tilde{q}$, $\tilde{S}_r(\tilde{q})$ decays as $\tilde{q}^{-3}$, as
indicated by the continuous line.

Figure \ref{ener.vac} shows a log-log plot of the time evolution of
$\sigma_r(t)$, $\langle\Delta E(t)\rangle$ and the best fits of the power
laws defined by eq.~(\ref{plr}). The resulting $x$ and $y$ exponents, are
listed in Table \ref{tabvac}, and plotted in front of $1/L$ in
Fig.~\ref{exp.vac}. Extrapolations to $L \rightarrow \infty$ render $x\simeq
y\simeq 0.36$ for $T_q= 1.0J/k_B$ and $x \simeq y \simeq 0.44$ for
$T_q=1.5J/k_B$.

The profiles of the transverse scan of the structure factor are plotted in
Fig.~\ref{fact.vac}. The same data is presented in scaled form in
Fig.~\ref{facesct.vac}. Notice that the scaling is again satisfied at $q=0$.
Scaling also holds for data corresponding to different system sizes. The
continuous lines show fits of lorentzian functions [eq.~(\ref{fitt})] with
$\alpha=1$ and $\sigma=1$. The behavior of the tail is
$\tilde{S}_t(\tilde{q}) \propto \tilde{q}^{-2}$, as we obtained for the
atom-atom exchange case.

The time evolution of $\sigma_t(t)$ can be seen in Fig.~\ref{elengtht.vac}
and the fitted exponents according to eq.~(\ref{plt}) are plotted in
Fig.~\ref{exp.vac} and listed in Table \ref{tabvac}. The extrapolations to $L
\rightarrow \infty$ render $z \simeq 0.34$ for $T_q= 1.0J/k_B$ and $z \simeq
0.44$ for $T_q= 1.5J/k_B$. Finally, the time evolution of the order parameter
is shown in Fig.~\ref{smax.vac}.

\section{Discussion}
\label{Discussion}

The analysis of the growth exponents for finite $L$ reveals that,
independently of the exchange mechanism and temperature, the exponents $x$
and $y$ corresponding to $\langle \Delta E(t) \rangle $ and $\sigma_r(t)$
respectively, coincide within errors bars: the obtained numerical value is,
at low temperature, lower than the Allen-Cahn growth exponent $1/2$;
nevertheless it raises towards such value when approaching the order-disorder
transition temperature. The value $y=1/2$ has been obtained experimentally
\cite{Shannon92} with high accuracy at quenching temperatures $T_q$ ranging
from $0.96T_0$ to $0.99T_0$ (our greatest simulated value is $T_q= 0.83T_0$).
Concerning the exponent $z$, characterizing the growth of $\sigma_t$, we
obtain $z \simeq x \simeq y$ for the case of vacancy-atom exchange mechanism
and $z < x \simeq y$ for the case of atom-atom exchange mechanism. In the
later case, the difference becomes more important with decreasing
temperature. We expect that, even in this case, $z$ will reach the value
$1/2$ when $T_q$ approaches $T_0$. Experimentally, no clear differences have
been observed between $y$ and $z$. This could be due either to the fact that
the relevant physical mechanism for the growth is the vacancy-atom exchange,
or that the experiments are performed at temperatures too close to $T_0$.
Nevertheless our simulation results, at low temperatures, suggest that the
evolution of $\sigma_t$ cannot be described by the standard Allen-Cahn growth
law. Additional physical considerations are needed in order to describe the
evolution  of such flat interfaces in the case of atom-atom exchange
mechanism.

   From the values of the exponents corresponding to different system sizes we
have performed a finite size analysis following the method proposed in
Ref.~\cite{Frontera94}. This analysis assumes a first order correction to the
growth law according to:

\be
R(t) \sim t^n \left (1- \frac{b}{t} \right) ,
\ee
where $n$ stands for any of the growth exponents $x$, $y$ or $z$.
This implies that the finite size dependence of the exponent $n_L$
follows:

\be
n_L= n_{\infty} +b L^{-1/n_{\infty}} .
\ee

Fits of this equation to the obtained exponents are plotted (solid lines) in
Figs.~\ref{exp.sec} and \ref{exp.vac}. The resulting values of $n_{\infty}$
are shown in Tables \ref{tabsec} and \ref{tabvac}. These extrapolated values
confirm the points discussed  in the previous paragraph. The values of the
first order correction coefficient $b$ are shown in Table \ref{coefb}. For
the case of atom-atom exchange mechanism the sign of the coefficients $b$
corresponding to the exponents $z$ are negative reinforcing the suggestion
that the standard Allen-Cahn growth law does not hold in this case. It is
also interesting to remark that, in general, the coefficient $b$ is greater
for the vacancy-atom exchange mechanism than for the atom-atom exchange one.
This confirms that in the former case the exponents exhibit a greater
dependence with finite size. This is in agreement with previous results
reported for 2-d square lattices \cite{Vives93b}.

Concerning the scaling of the structure factor, it is worth noting that it
holds in a broad range of time for all the studied cases indicating that the
domain growth regime, in which the evolution of the system is governed by two
characteristic lengths proportional, respectively, to $\sigma_r^{-1}$ and
$\sigma_t^{-1}$, is clearly reached in our simulations. Whether or not both
lengths obey the same growth law can be detected by the scaling of the
structure factor at $q=0$. At this $q$-position, scaling holds only if
$\sigma_r(t) \propto \sigma_t(t)$. For the case of vacancy-atom exchange
mechanism scaling at $q=0$ is definitively satisfied, indicating that a
single growth law governs the evolution of the system consistently with an
unique value of the growth exponents ($x \simeq y \simeq z$). Contrarily, for
the case of atom-atom exchange mechanism, a lack of scaling at $q=0$ can be
clearly seen in Figs.~\ref{facescr.sec} and \ref{facesct.sec} (compare with
Figs.~\ref{facescr.vac} and \ref{facesct.vac} at $q=0$). This indicates that
the two lengths are needed to characterize the evolution of the system in
this case.

\begin{figure}

\centerline{\hbox{
\psfig{figure=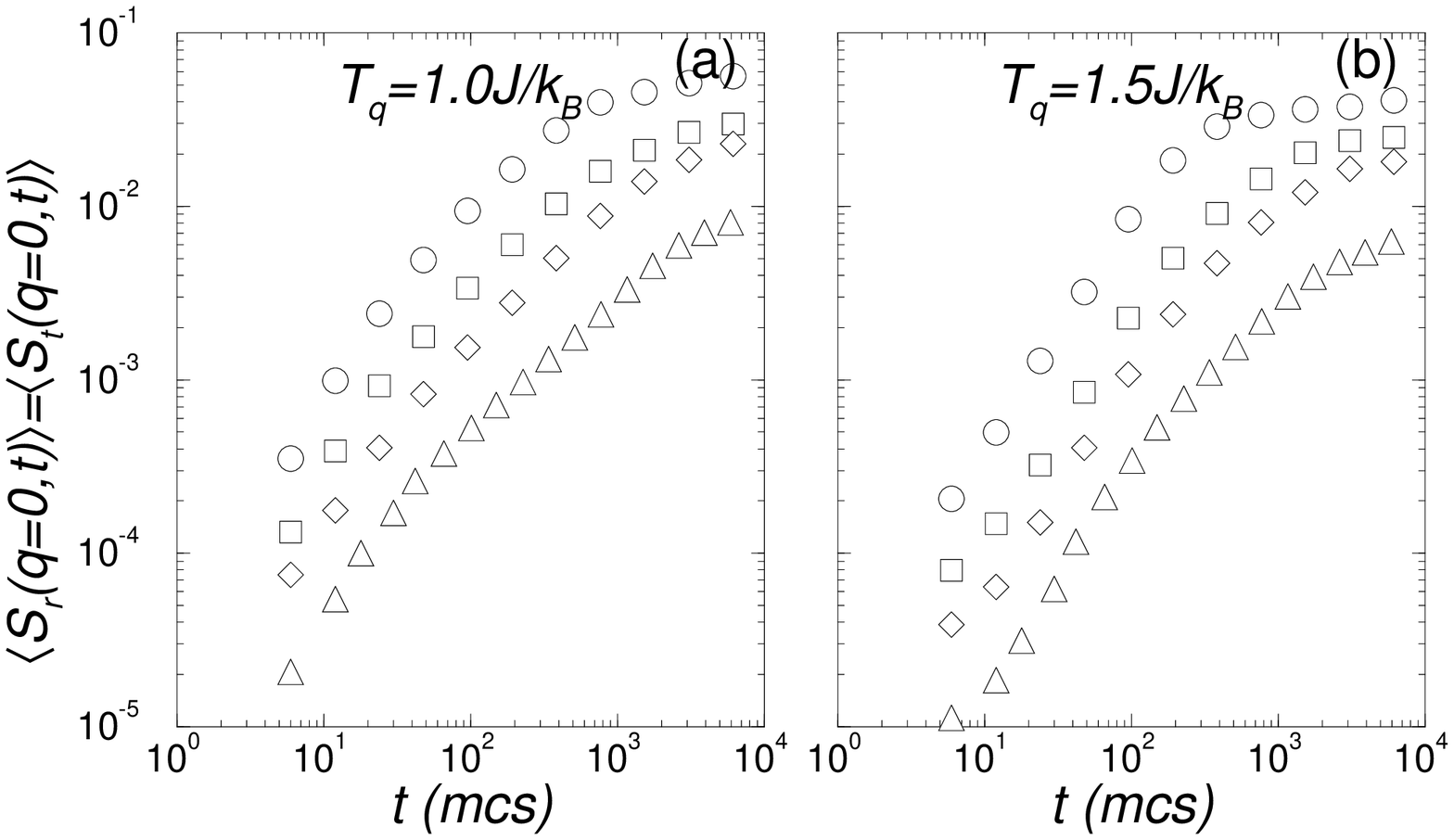,height=305pt,width=400pt}} }

\caption{Log-log plot of the value of the structure factor at the
superstructure peak $q = 0$ for different system sizes ($L=20$
($\circ $), $L=28$ ($\Box$), $L=36$ ($\Diamond $), $L=64$ ($\triangle$) and
different quenching temperatures with the atom-atom exchange mechanism. The
curves corresponding to different $L$ have been vertically shifted in order
to clarify the picture.} \label{smax.sec}

\end{figure}

\begin{figure}

\centerline{\hbox{
\psfig{figure=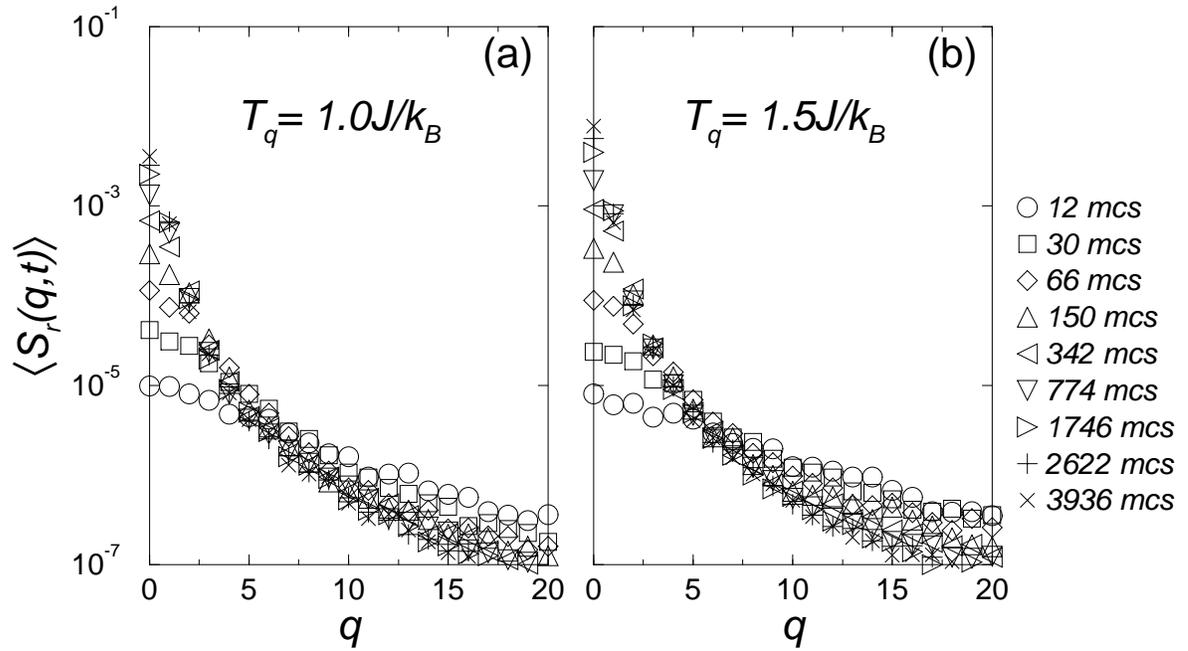,height=305pt,width=400pt}} }

\caption{Linear-log plot of the averaged radial scan of the structure factor
at different times and quenching temperatures $T_q= 1.0 J/k_B$ (a) and
$T_q=1.5J/k_B$ (b) for the vacancy-atom exchange mechanism. Both figures
correspond to systems of linear size $L=64$.}

\label{facr.vac}
\end{figure}

\begin{figure}

\centerline{\hbox{
\psfig{figure=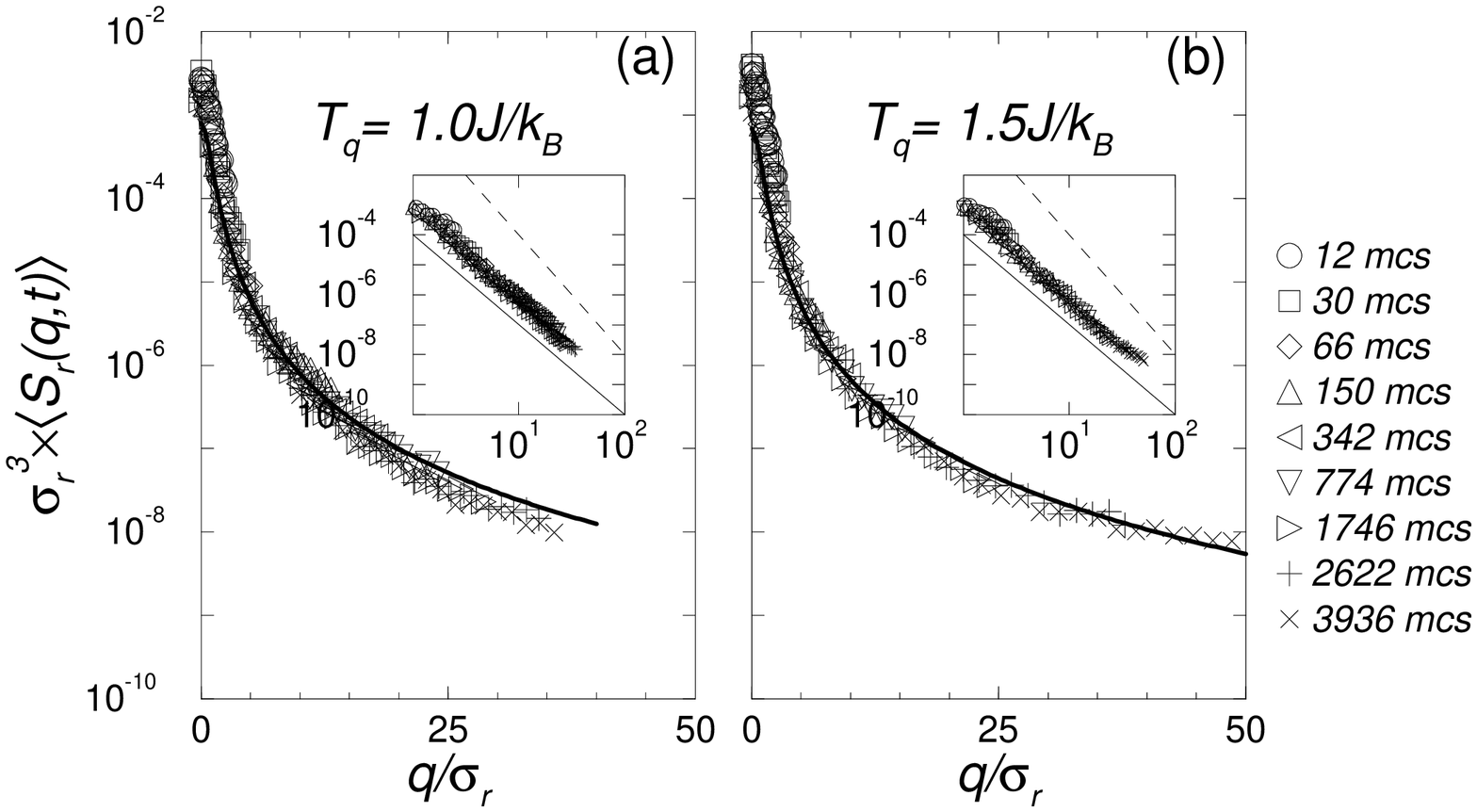,height=305pt,width=400pt}} }

\caption{Linear-log and log-log (insets) plots of the scaled radial scan
of the structure factor at different times and quenching temperatures $T_q=
1.0 J/k_B$ (a) and $T_q=1.5J/k_B$ (b) for the vacancy-atom exchange
mechanism. Data correspond to systems of linear size $L=64$. The solid thick
line corresponds to a fit of expression (3.7) with $\alpha=1.5$ and
$\sigma=1$. The dashed lines in the insets show the Porod's law
$\tilde{S}_r(\tilde{q})\sim \tilde{q}^{-4}$, while solid lines show the
slope of $\tilde{q}^{-3}$.}

\label{facescr.vac}
\end{figure}

\begin{figure}

\centerline{\hbox{
\psfig{figure=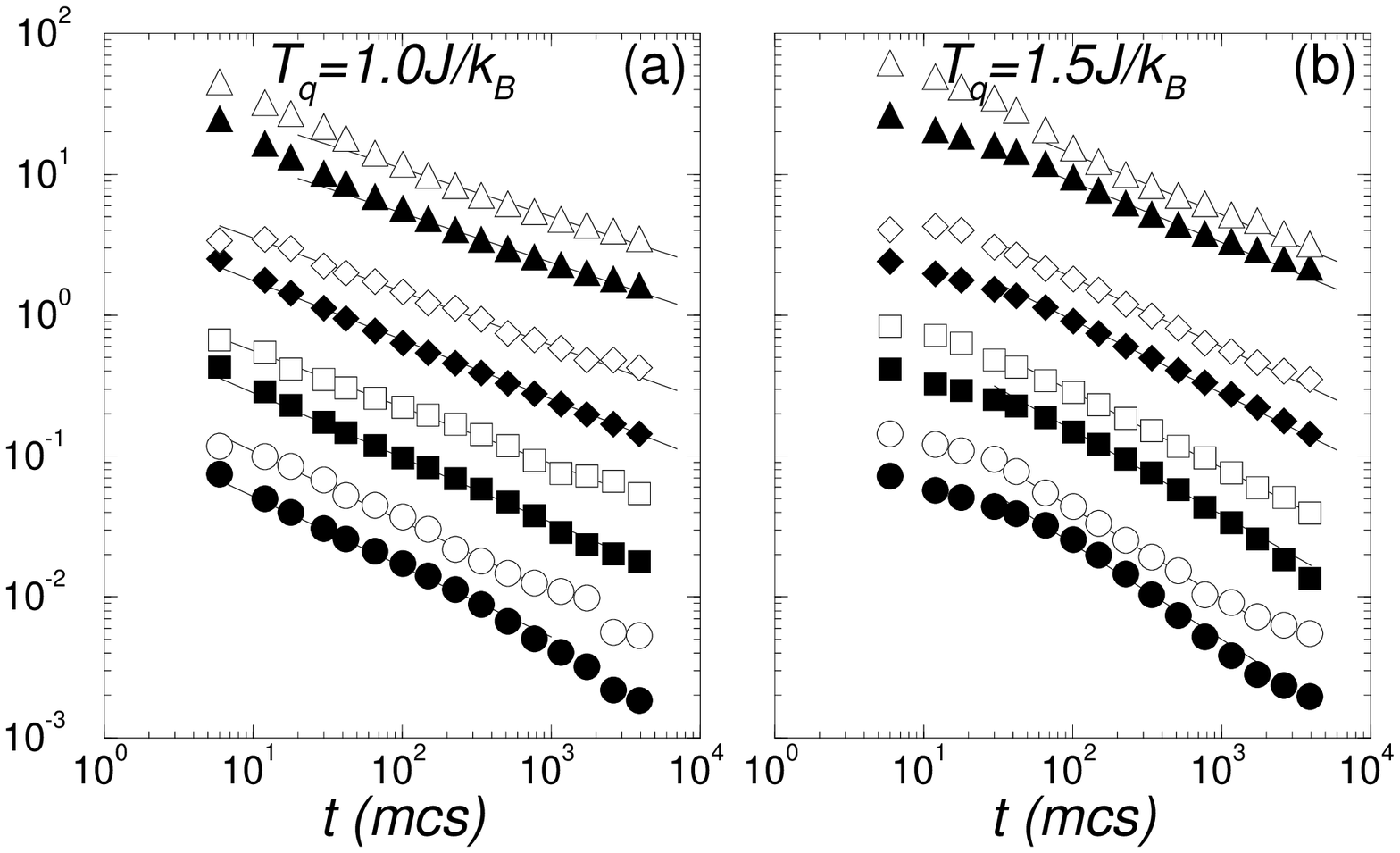,height=305pt,width=400pt}} }

\caption{Time evolution, in the case of vacancy-atom exchange mechanism, of
$\langle\Delta E(t)\rangle$ (filled symbols) and of $\sigma_r(t)$ (open
symbols), for $T_q= 1.0J/k_B$ (a) and $T_q=1.5J/k_B$ (b) for systems of
$L=20$ (circles), $L=28$ (squares), $L=36$ (diamonds) and $L=64$ (triangles).
The solid lines are the best power-law fits. Data corresponding to different
sizes have been vertically shifted to clarify the picture.}

\label{ener.vac}
\end{figure}

\begin{figure}

\centerline{\hbox{
\psfig{figure=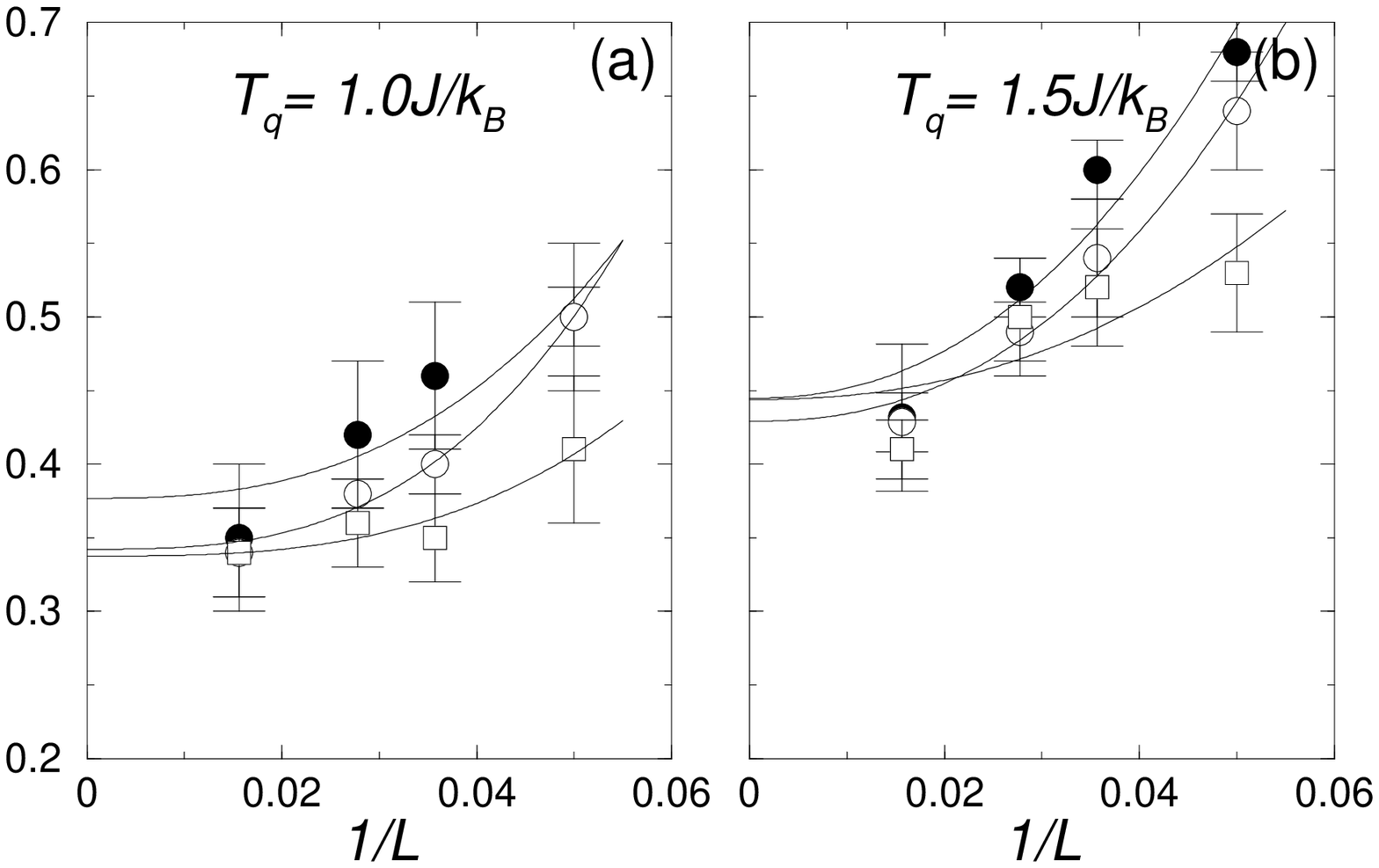,height=305pt,width=400pt}} }

\caption{Exponents $x$ ($\bullet$) from the excess energy per site ($\Delta
E$), $y$ ($\circ$) from the second moment of the radial scan of the
superstructure peak $\sigma_r$ and $z$ ($\Box$) from the fitted amplitude of
the transverse scan of the superstructure peak $\sigma_t$, as a function of
$1/L$ with the vacancy-atom exchange mechanism. Data corresponds to
temperature $T_q=1.0J/k_B$ (a) and $T_q=1.5J/k_B$ (b). Solid lines show the
best fits of eq. (5.2).}

\label{exp.vac}
\end{figure}

\begin{figure}

\centerline{\hbox{
\psfig{figure=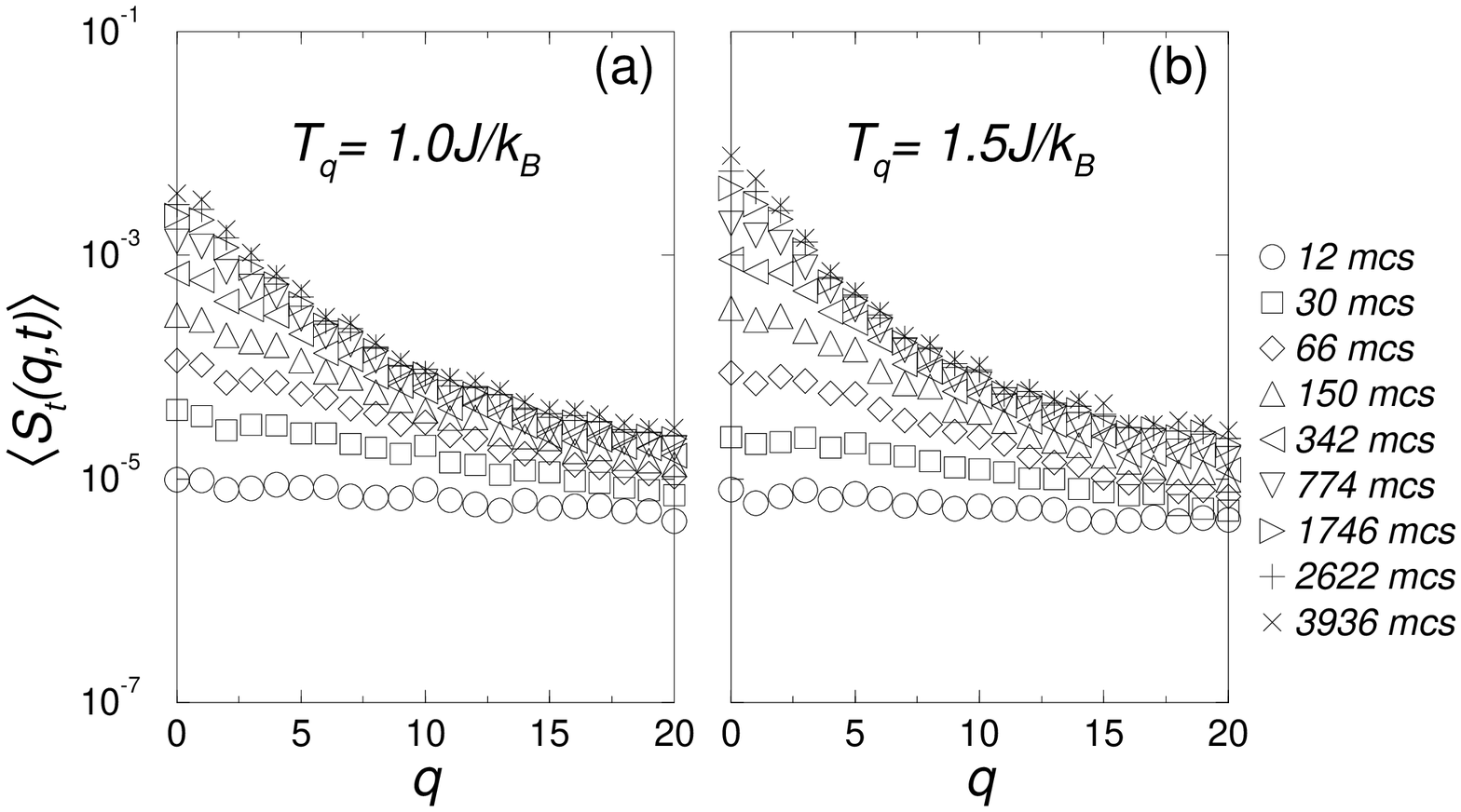,height=305pt,width=400pt}} }

\caption{Linear-log plot of the averaged transverse scan of the structure
factor at different times and quenching temperatures $T_q= 1.0 J/k_B$ (a)
and $T_q=1.5J/k_B$ (b) with the vacancy-atom exchange mechanism. Data
correspond to systems of linear size $L=64$.}

\label{fact.vac}
\end{figure}

\begin{figure}

\centerline{\hbox{
\psfig{figure=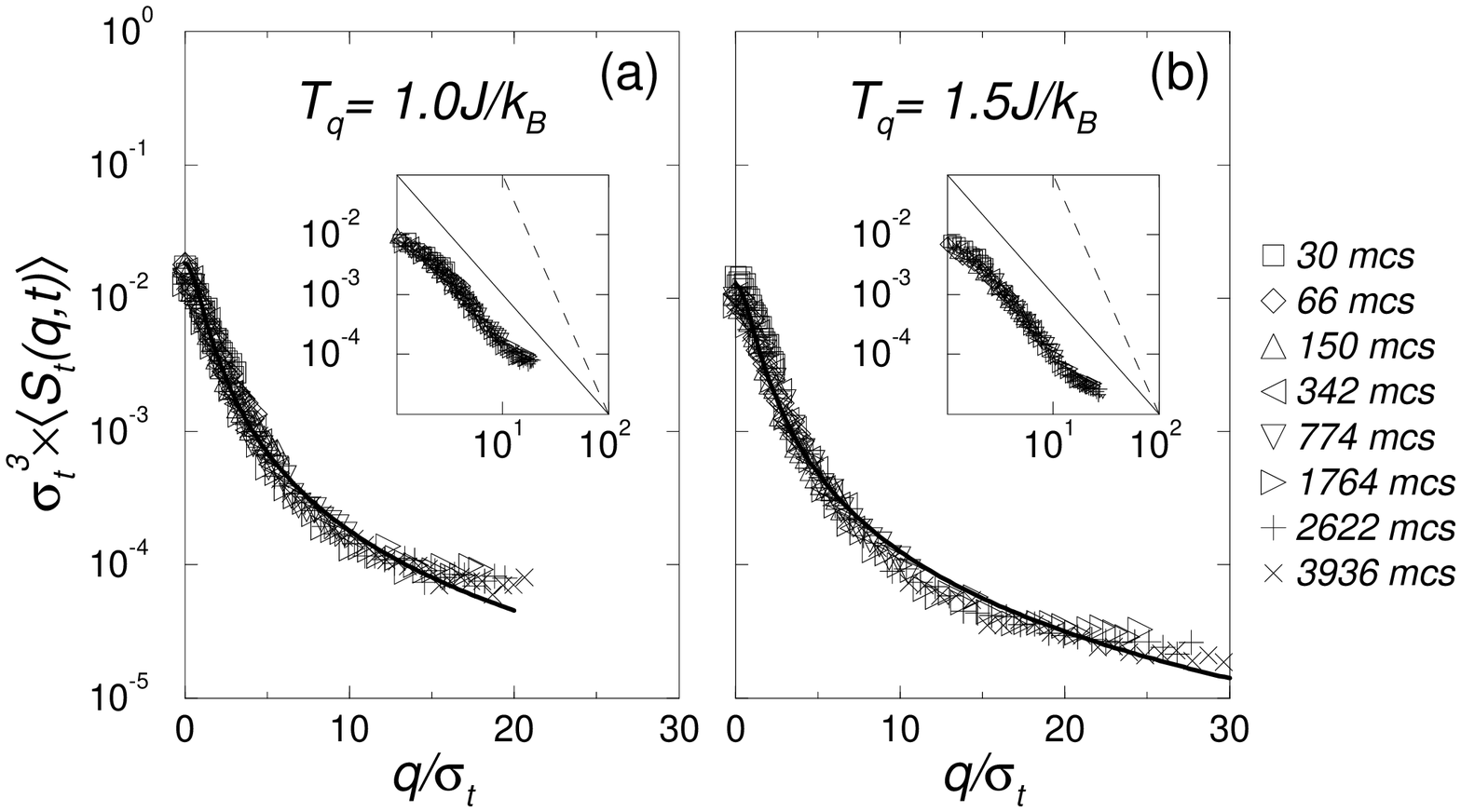,height=305pt,width=400pt}} }

\caption{Linear-log and log-log (insets) plots of the scaled transverse scan
of the structure factor at different times and quenching temperatures $T_q=
1.0 J/k_B$ (a) and $T_q=1.5J/k_B$ (b) with vacancy-atom exchange mechanism.
Data correspond to systems of linear size $L=64$. The solid thick lines
correspond to fits of expression (3.7) with $\alpha=1$ and $\sigma=1$. The
dashed lines in the insets show the Porod's law $\tilde{S}(\tilde{q})\sim
\tilde{q}^{-4}$, while the solid lines show the slope of $\tilde{q}^{-2}$.}

\label{facesct.vac}
\end{figure}

\begin{figure}

\centerline{\hbox{
\psfig{figure=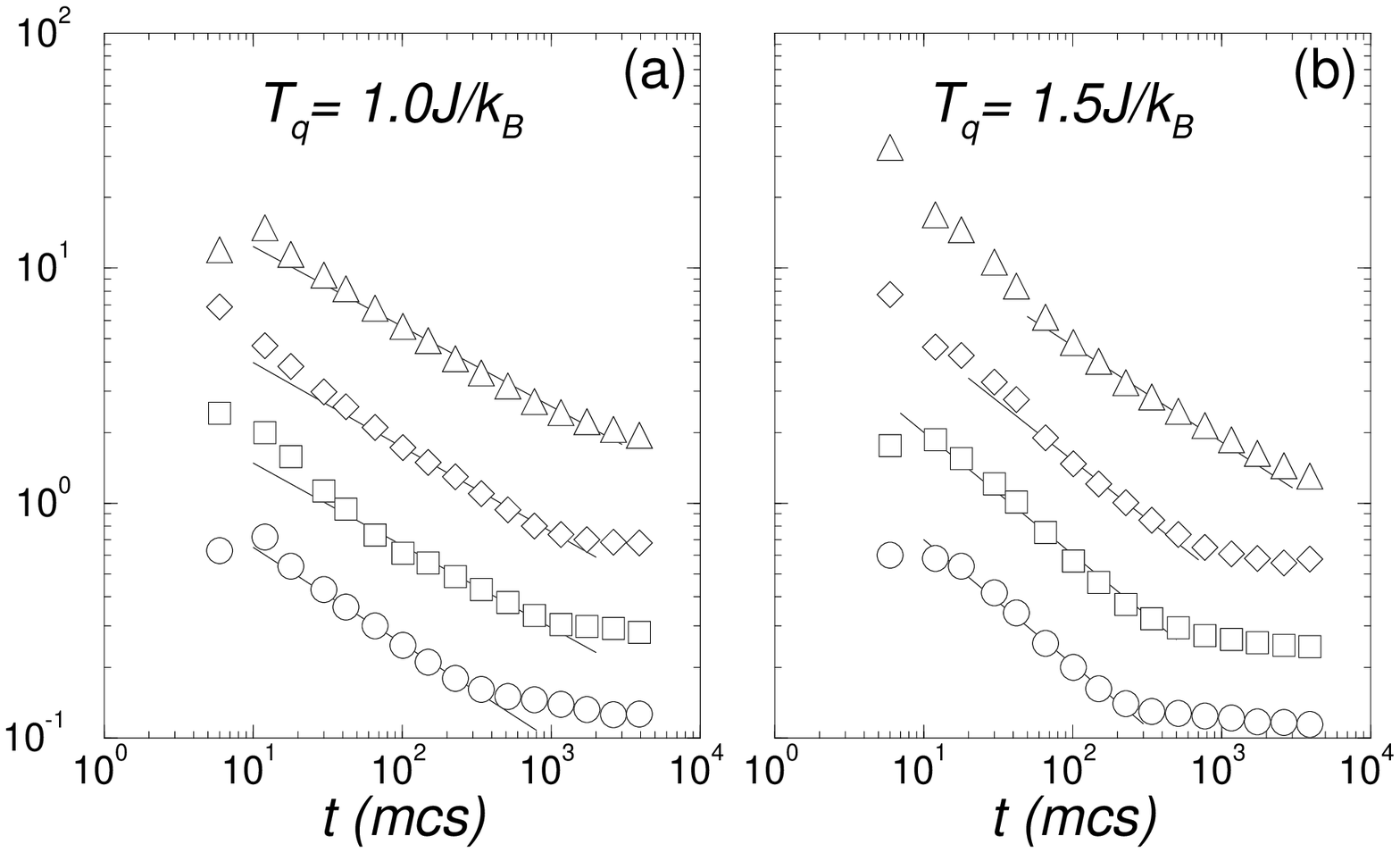,height=305pt,width=400pt}} }

\caption{Log-log plot of the mean distance between type-1 walls for $L=20$
($\circ $), $L=28$ ($\Box$), $L=36$ ($\Diamond $), $L=64$ ($\triangle$) with
vacancy-atom exchange mechanism. Data correspond to quenching temperatures
$T_q= 1.0J/k_B$ (a) and $T_q=1.5J/k_B$ (b). Solid lines show the best
power-law fits.}

\label{elengtht.vac}
\end{figure}

\begin{figure}

\centerline{\hbox{
\psfig{figure=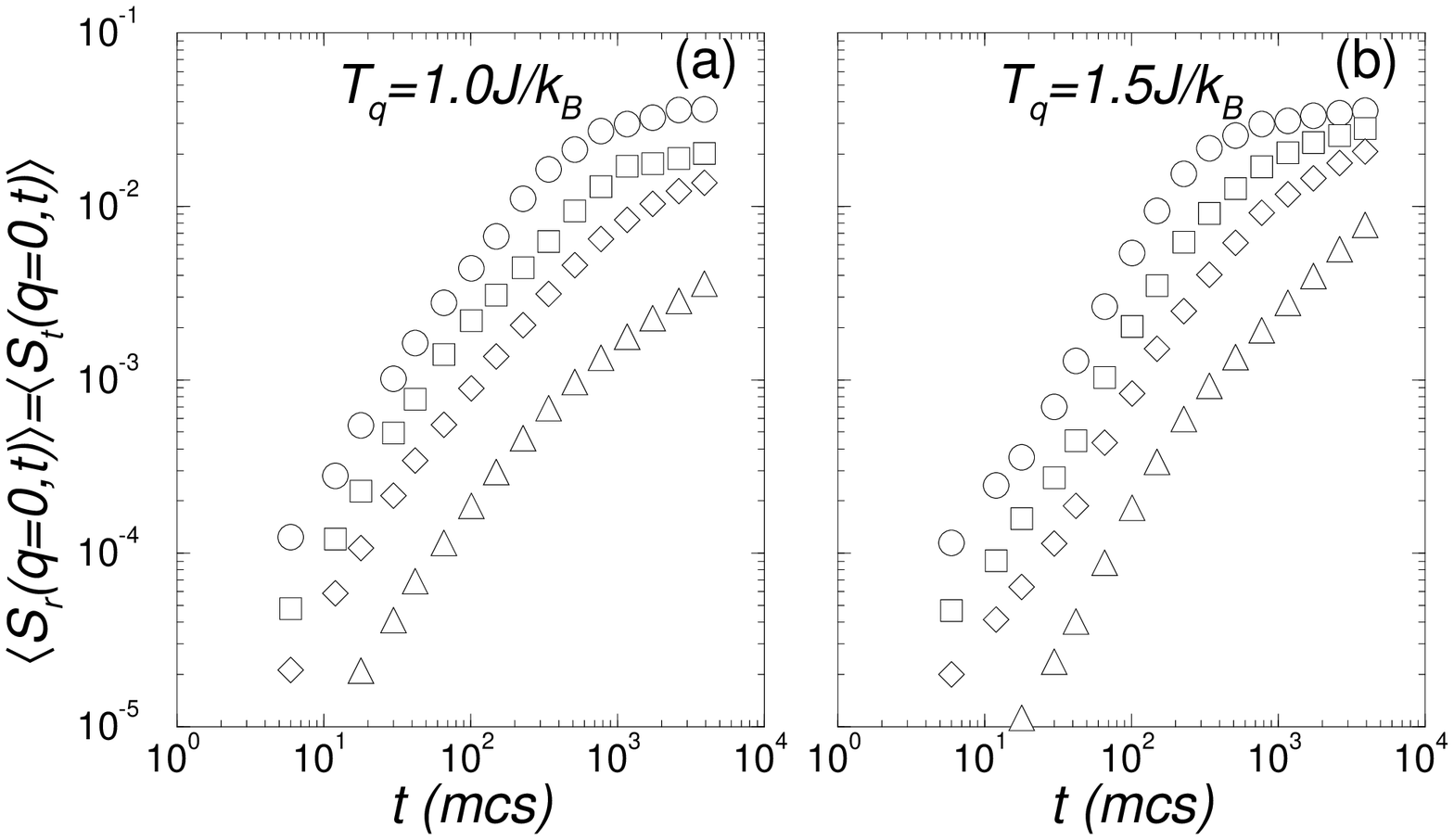,height=305pt,width=400pt}} }

\caption{Log-log plot of the value of the structure factor at the
superstructure peak $q = 0$ for different system sizes ($L=20$
($\circ $), $L=28$ ($\Box$), $L=36$ ($\Diamond $), $L=64$ ($\triangle$) and
different quenching temperatures with the vacancy-atom exchange mechanism.
The curves corresponding to different $L$ have been vertically shifted in
order to clarify the picture.}

\label{smax.vac}
\end{figure}

The time evolution of the order parameter [$\langle S(q=0) \rangle $] shown
in Figs.~\ref{smax.sec} and \ref{smax.vac} is in qualitative good agreement
with the experimental results for $Cu_3Au$ (see Fig.~18 in
Ref.~\cite{Shannon92}). Even the existence of a possible delay time due to an
incubation period for nucleation, that has been found experimentally, can be
observed as an inflexion point in our curves. In our model the disordered
phase is metastable down to zero temperature due to frustration effects
\cite{Binder80}. Therefore the evolution, in our simulations, initiates via
nucleation, as expected \cite{Ludwig88,Gaulin90} for $Cu_3Au$ in the
experimentally studied temperature range \cite{Shannon92}. Also in agreement
with experiments, the delay time increases with increasing $T_q$. The
experimentalists \cite{Shannon92} suggest an explanation for this delay based
on the influence of the elastic energy in the nucleation process. Despite not
containing elastic effects our model reproduces such results, indicating that
a full explanation cannot rely only on elasticity arguments. Moreover,
comparing Figs. \ref{smax.sec}(b) and \ref{smax.vac}(b) it seems that the
delay is more important in the case of the vacancy mechanism. This
problem will be studied in a future work.

\begin{figure}

\centerline{\hbox{
\psfig{figure=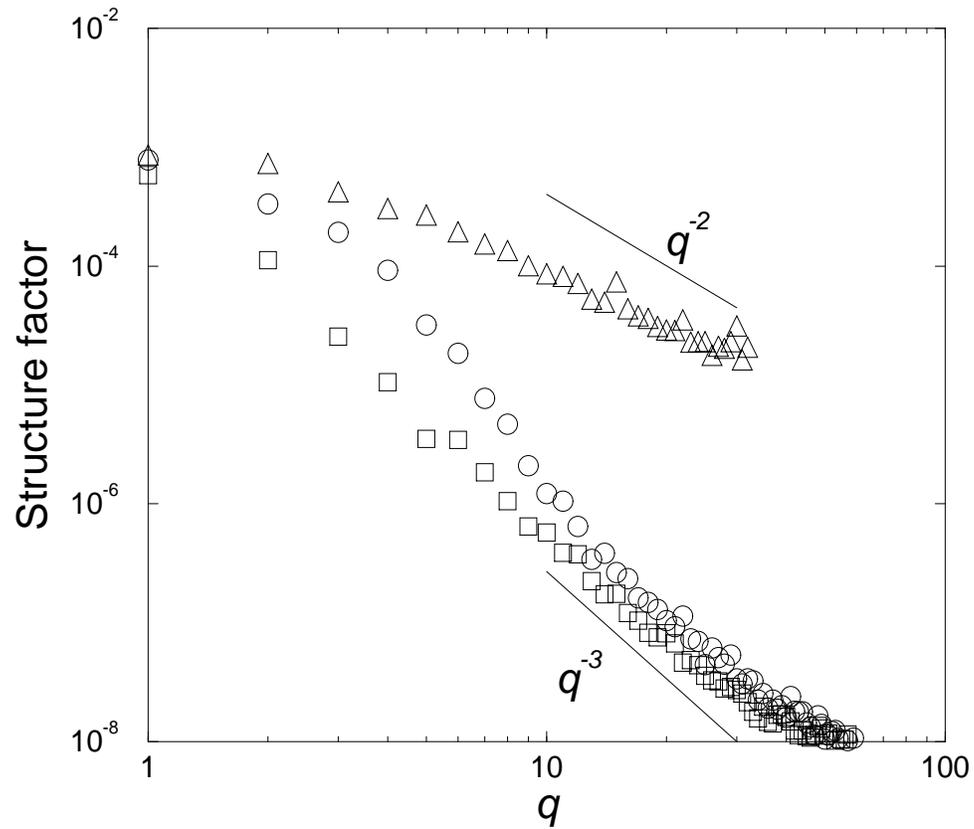,height=335pt,width=400pt}} }

\caption{Log-log plot of the structure factor decay in the radial ($\Box$),
transversal($\triangle$) and diagonal ($\circ$) directions. Data, obtained
with the atom-atom exchange mechanism, correspond to $t= 1998 mcs$, $T_q=
1.0J/k_B$ and $L=64$.}

\label{dporod}
\end{figure}

The behaviour of the tail of the structure factor along the radial and the
transverse directions is markedly different. This can be clearly seen by
comparing the insets of Figs.~\ref{facescr.sec} and \ref{facescr.vac}
corresponding to the  radial scans and  those in Figs.~\ref{facesct.sec} and
\ref{facesct.vac} corresponding to the transverse scans. For such anisotropic
peaks, the Porod's Law \cite{Porod82}, $\langle S(q) \rangle \sim
q^{-(d+1)}$, ($d+1=4$ in our case) is not expected to be satisfied.
Independently of the dynamical exchange mechanism we have found that for
large values of $q$, $\langle S_r(q)\rangle \sim q^{-3}$ and $\langle
S_t(q)\rangle \sim q^{-2}$. This is in agreement with the fact that the best
lorentzian fits are obtained for $\alpha=3/2$ and $\alpha= 1$ for radial and
transverse scans respectively, as explained in section \ref{MCDetails}. We
have also studied the dependence of $\langle S(q)\rangle $ for large $q$
along the diagonal direction [$\vec{k}= (1,1,0)+ \frac{1}{L}(q,q,0)$]. Figure
\ref{dporod} compares the decay along this direction with the transverse and
radial ones. For the diagonal direction, results are consistent with a
$q^{-3}$ decay. This reveals the singular character of the behaviour along
the transverse direction. It is interesting to remark that the present
simulation results are not in agreement with the theory for ordering dynamics
in $Cu_3Au$ proposed by Lai \cite{Lai90}. In that theory it is found that the
Porod's Law is satisfied for both radial and transverse scans. This important
difference between Lai's theory and the present simulation probably arises
from the fact that in our model type-1 walls have zero excess energy.

Concerning the previous studies of the order-disorder dynamics by means of
the vacancy-atom exchange mechanism some important differences must be
pointed out. Firstly, for two dimensional square \cite{Frontera93} and bcc
lattices \cite{Frontera94} vacancies are known to accelerate the growth
leading, at low temperatures, to exponents $x > 1/2$. This is due to the fact
that the vacancy prefers to  move along the interfaces, increasing the number
of accepted exchanges. Such acceleration only appears if the vacancy is
allowed to jump to next-nearest neighbor (n.n.n.) positions. Such jumps
prevent the vacancy to be trapped in the ordered regions. When only jumps to
n.n. positions are permitted, the growth law becomes logarithmic. Although in
the present simulations on fcc lattices we only allow vacancy jumps to n.n.
positions, trapping does not appear since for such fcc lattices the ordered
regions can be crossed without energy barriers. We have tested, by performing
a number of simulations allowing the vacancy to jump to n.n.n. positions,
that the growth exponents do not change significantly. Secondly, for the
present simulations the acceleration of the domain growth process compared to
the atom-atom exchange mechanism does not appear. This is due to the fact
that the tendency for the vacancy to sit on the interfaces is quite low in
this case: some of the interfaces are non energetic and, among the energetic
ones, only those with an excess of the majoritary specie represent an energy
gain for the vacancy compared to the bulk. Consequently the vacancy path in
this case is much more homogeneous (close to a random walk) than for the case
of a bcc lattice (close to a self-avoiding walk) \cite{Frontera94}.

Finally, there is still a very important question to be answered: Why, in the
case of vacancy-atom exchange mechanism, the growth can be described by a
single length whereas two different lengths are needed in the atom-atom
exchange mechanism case? Actually, for the case of atom-atom exchange
mechanism the need of  two characteristic lengths is expected, as has been
found in several studies of anisotropic growth. For instance, in a
two-dimensional model for a martensitic transition \cite{Castan89}, the
existence of two relevant lengths arises as a consequence of the coexistence
of flat and curved interfaces, that can only evolve in a hierarchical way:
the curved interfaces have to wait for the planar interfaces to disappear
before they can decrease in length. It is worth mentioning that in such model
the anisotropy is explicitly introduced in the hamiltonian, while in the
present case anisotropy appears due to the topology of the fcc lattice. It is
found, in agreement with our result using the atom-atom (at low $T_q$)
exchange mechanism, that the mean distance between planar interfaces grows as
$t^{1/4}$ and that the mean distance between curved interfaces grows as
$t^{1/2}$. Also, in a model for island growth \cite{Vinals85}, the
anisotropic growth appears as a consequence of the coexistence of two kinds
of interfaces. The surprising result is that the vacancy-atom exchange
mechanism destroys the expected anisotropic growth. We cannot provide a
definitive explanation for this question at present; nevertheless, we believe
that the sequentiality of the vacancy path modifies the hierarchical motion
of the planar and curved interfaces existing in the system.


\section{Summary and conclusions}
\label{Summary}

In this paper extensive Monte Carlo simulations of $L1_2$ ordering kinetics
on a fcc $A_3B$ binary alloy are presented. We have focused on the study of
the evolution of the superstructure peak and the excess energy. We have
followed the growth up to $10^4 \; mcs$ in systems with linear size ranging
from $L=20$ to $L=64$. Two different dynamics have been used: first the
standard Kawasaki atom-atom exchange mechanism and second, the more realistic
vacancy-atom exchange mechanism. In this last case a very small concentration
of vacancies is introduced in the system. Finite size scaling techniques have
been used in order to extrapolate to $L\rightarrow \infty$ the growth
exponents evaluated at finite $L$. We have obtained that finite-size effects
are more important in the case of the vacancy-atom exchange mechanism. For
the atom-atom exchange mechanism we have found evidences of anisotropic
growth: the width of the superstructure peak in the transverse direction
evolves according to $\sigma_t\sim t^{-z}$ with $z$ smaller than the exponent
$y$ characterizing the evolution of the width in the radial direction
$\sigma_r\sim t^{-y}$. Such anisotropy arises from the hierarchical interface
motion of the two different domain walls. As proposed in a recent theory
\cite{Castan89}, this leads to $y=1/2$ and $z=1/4$ in rather good agreement
with our experimental results. Contrarily, for the vacancy-atom mechanism,
the two relevant lengths $\sigma_t^{-1}$ and $\sigma_r^{-1}$ evolve following
the same power law. The disappearance of the anisotropic growth with the
vacancy-atom exchange mechanism may arise from the fact that the
sequentiality of the vacancy path destroys the hierarchical evolution of the
two different interfaces existing in the system.

The effect of the temperature has also been studied. For both mechanism the
exponents $y$ and $z$ tend to 1/2 when the transition temperature is
approached from below. To our knowledge, all the experiments have been
performed at $T_q \sim T_0$, so that no conclusive results about the relevant
ordering mechanisms can be deduced. The results in this paper suggest that
more experimental studies of the ordering dynamics, specially at low
temperatures, are desirable in order to clarify the existence of such
anisotropic growth and which is the relevant mechanism for ordering.
Synchrotron radiation facilities seem to be very promising for the obtention
of the needful structure factor maps in the late time regime.

\acknowledgements

We acknowledge the Comisi\'on interministerial de Ciencia y tecnolog\'{\i}a
(CICyT) for financial support (Project No. MAT95-504) and the Fundaci\'o
Catalana per a la Recerca (FCR) and the Centre de Supercomputaci\'o de
Catalunya (CESCA) for computational facilities. C.F. also acknowledges
financial support from the Comissionat per a Universitats i Recerca
(Generalitat de Catalunya).



\begin{table}[h]

\caption{Directions of the non-energetic type-1 walls between possible pairs
of neighboring domains.}

\begin{tabular}{c|c|c|c|c|}

\begin{tabular}{c} \\ $(\Psi_1,\Psi_2,\Psi_3)$ \end{tabular}
& \begin{tabular}{c}
$\alpha$ \\ $(-1,-1,-1)$
\end{tabular}
& \begin{tabular}{c}
$\beta$ \\ $(-1,1,1)$
\end{tabular}
& \begin{tabular}{c}
$\gamma$ \\ $(1,-1,1)$
\end{tabular}
& \begin{tabular}{c}
$\delta$  \\ $(1,1,-1)$
\end{tabular}
\\ \hline
$\alpha$ & & $[100]$ & $[010]$ & $[001]$ \\ \hline
$\beta$ & $[100]$ & & $[001]$ & $[010]$ \\ \hline
$\gamma$ & $[010]$ & $[001]$ & & $[100]$ \\ \hline
$\delta$ & $[001]$ & $[010]$ & $[100]$ &

\end{tabular}

\label{Direc}

\end{table}

\begin{table}[h]

\caption{Fitted exponents of the power-law evolution of the excess energy per
site ($\Delta E$), of the second moment of the radial scan ($\sigma_r$) and
of the fitted amplitude of the transverse scan ($\sigma_t$). Data correspond
to simulations with the atom-atom exchange mechanism.}

\label{tabsec}

\begin{tabular}{c|c|c|c|c|c|}

\begin{tabular}{lr} & $L$ \\ $k_B T_q/J$ & \end{tabular} & 20 & 28 & 36 & 64
& $L\rightarrow \infty$ \\
\hline
1.0 \begin{tabular}{c} $\Delta E$ \\ $\sigma_r$ \\$\sigma_t$ \end{tabular} &
\begin{tabular}{c} $0.46 \pm 0.02$ \\ $0.44 \pm 0.02$ \\
$0.25 \pm 0.03$ \end{tabular} &
\begin{tabular}{c} $0.41 \pm 0.03$ \\ $0.42 \pm 0.03$ \\
$0.25 \pm 0.04$ \end{tabular} &
\begin{tabular}{c} $0.43 \pm 0.03$ \\ $0.42 \pm 0.02$ \\
$0.29 \pm 0.02$ \end{tabular} &
\begin{tabular}{c} $0.40 \pm 0.02$ \\ $0.41 \pm 0.02$ \\
$0.25 \pm 0.03$ \end{tabular} &
\begin{tabular}{c} $0.40 \pm 0.02$ \\ $0.41 \pm 0.02$ \\
$0.26 \pm 0.02$ \end{tabular} \\
\hline
1.5 \begin{tabular}{c} $\Delta E$ \\ $\sigma_r$ \\$\sigma_t$ \end{tabular} &
\begin{tabular}{c} $0.50 \pm 0.06$ \\ $0.50 \pm 0.05$ \\
$0.41 \pm 0.05$ \end{tabular} &
\begin{tabular}{c} $0.49 \pm 0.03$ \\ $0.48 \pm 0.03$ \\
$0.45 \pm 0.04$ \end{tabular} &
\begin{tabular}{c} $0.50 \pm 0.02$ \\ $0.48 \pm 0.03$ \\
$0.45 \pm 0.06$ \end{tabular} &
\begin{tabular}{c} $0.44 \pm 0.03$ \\ $0.39 \pm 0.02$ \\
$0.47 \pm 0.05$ \end{tabular} &
\begin{tabular}{c} $0.46 \pm 0.03$ \\ $0.42 \pm 0.02$ \\
$0.47 \pm 0.02$ \end{tabular}
\end{tabular}

\end{table}

\begin{table}[h]

\caption{Fitted exponents of the power-law evolution of the excess energy per
site ($\Delta E$), of the second moment of the radial scan ($\sigma$), and of
the fitted amplitude of the transverse scan ($\sigma_t$). Data correspond to
simulations with the vacancy-atom exchange mechanism.}

\label{tabvac}

\begin{tabular}{c|c|c|c|c|c|}

\begin{tabular}{lr} & $L$ \\ $k_B T_q/J$ & \end{tabular} & 20 & 28 & 36 & 64
& $L\rightarrow \infty$ \\
\hline
1.0 \begin{tabular}{c} $\Delta E$ \\ $\sigma_r$ \\ $\sigma_t$
\end{tabular} &
\begin{tabular}{c} $0.50 \pm 0.02$ \\ $0.50 \pm 0.02$ \\
$0.41 \pm 0.03$ \end{tabular} &
\begin{tabular}{c} $0.46 \pm 0.03$ \\ $0.40 \pm 0.03$ \\
$0.35 \pm 0.04$ \end{tabular} &
\begin{tabular}{c} $0.42 \pm 0.03$ \\ $0.38 \pm 0.02$ \\
$0.36 \pm 0.02$ \end{tabular} &
\begin{tabular}{c} $0.35 \pm 0.02$ \\ $0.34 \pm 0.02$ \\
$0.34 \pm 0.03$ \end{tabular} &
\begin{tabular}{c} $0.38 \pm 0.02$ \\ $0.34 \pm 0.03$ \\
$0.34 \pm 0.02$ \end{tabular} \\
\hline
1.5 \begin{tabular}{c} $\Delta E$ \\ $\sigma_r$ \\ $\sigma_t$
\end{tabular} &
\begin{tabular}{c} $0.68 \pm 0.05$ \\ $0.64 \pm 0.03$ \\
$0.53 \pm 0.05$ \end{tabular} &
\begin{tabular}{c} $0.60 \pm 0.05$ \\ $0.54 \pm 0.02$ \\
$0.52 \pm 0.04$ \end{tabular} &
\begin{tabular}{c} $0.52 \pm 0.04$ \\ $0.49 \pm 0.02$ \\
$0.50 \pm 0.06$ \end{tabular} &
\begin{tabular}{c} $0.43 \pm 0.05$ \\ $0.43 \pm 0.02$ \\
$0.41 \pm 0.05$ \end{tabular} &
\begin{tabular}{c} $0.44 \pm 0.02$ \\ $0.43 \pm 0.01$ \\
$0.44 \pm 0.04$ \end{tabular}

\end{tabular}

\end{table}

\begin{table}

\caption{Fitted coefficients $b$ of Eq. (5.2)}

\begin{tabular}{c|c|c|}
$k_B T_q/J$ & atom-atom mechanism & vacancy-atom mechanism\\
\hline
$1.0$
\begin{tabular}{c} $\Delta E$ \\ $\sigma_r$ \\ $\sigma_t$\end{tabular} &
\begin{tabular}{c} 98.7 \\ 45.4 \\ -1271.5 \end{tabular} &
\begin{tabular}{c} 386.1 \\ 1003.9 \\ 495.6 \end{tabular} \\
\hline
$1.5$
\begin{tabular}{c} $\Delta E$ \\ $\sigma_r$ \\ $\sigma_t$\end{tabular} &
\begin{tabular}{c} 31.2 \\ 106.2 \\ -34.7 \end{tabular} &
\begin{tabular}{c} 212.3 \\ 233.3 \\ 88.4 \end{tabular}

\end{tabular}

\label{coefb}

\end{table}

\end{document}